\documentclass[a4paper,11pt]{article}

\usepackage{jcappub} 

\usepackage[T1]{fontenc} 
\usepackage{aas_macros}
\usepackage{natbib}

\newtheorem{theorem}{Theorem}
\newtheorem{definition}{Definition}

\title{\boldmath A Study on the Baryon Acoustic Oscillation with Topological Data Analysis}

\author[a]{K.~T.~Kono,\note{Corresponding author.}}
\author[a,b]{T.~T.~Takeuchi,}
\author[a]{S.~Cooray,}
\author[c,a]{A.~J.~Nishizawa,}
\author[a]{K.~Murakami}

\affiliation[a]{Division of Particle and Astrophysical Science, Nagoya University,Furo-cho, Chikusa-ku, Nagoya 464--8602, Japan}
\affiliation[b]{The Research Center for Statistical Machine
Learning, the Institute of Statistical Mathematics, 10--3 Midori-cho, Tachikawa, Tokyo 190--8562, Japan}
\affiliation[c]{Institute for Advanced Research, Nagoya University, Furo-cho Chikusa-ku Nagoya 464--8601 Japan}  

\emailAdd{kono.kai@c.mbox.nagoya-u.ac.jp}
\emailAdd{takeuchi.tsutomu@g.mbox.nagoya-u.ac.jp}
\emailAdd{cooray@nagoya-u.jp}
\emailAdd{atsushi.nishizawa@iar.nagoya-u.ac.jp}
\emailAdd{murakami.koya@a.mbox.nagoya-u.ac.jp}

\abstract{
The scale of the acoustic oscillation of baryons at the baryon-photon decoupling is imprinted on the spatial distribution of galaxies in the Universe, known as the baryon acoustic oscillation (BAO). 
The BAO provides very useful information in the early Universe.  
The correlation functions and power spectrum are used as a central tool for the studies on the BAO analysis. 
In this work, we analyzed the spatial distribution of galaxies with a method from the topological data analysis (TDA), in order to detect and examine the BAO signal in the galaxy distribution.
The TDA provides a method to treat various types of "holes" in point set data, by constructing the persistent homology (PH) group from the geometric structure of data points and handling the topological information of the dataset. 
We can obtain the information on the size, position, and statistical significance of the holes in the data. 
A particularly strong point of the persistent homology is that it can classify the holes by their spatial dimension, i.e., a 0-dim separation, 1-dim loop, 2-dim shell, etc.  
We first analyzed the simulation datasets with and without the baryon physics to examine the performance of the PH method. 
We found that the PH is indeed able to detect the BAO signal among the large-scale structures in the Universe: simulation data with baryon physics present a prominent signal from the BAO, while data without baryon physics does not show this signal.  
In fact, a systematic method to test the significance of the signal detection from the PH diagram has not been established yet. 
Then, this examination can be a new reliable method to demonstrate the  performance of the PH analysis of the large-scale structure. 
Then, we applied the PH to a quasar sample at $z < 1.0$ from extended Baryon Oscillation
Spectroscopic Survey \cite{2016AJ....151...44D} in Sloan Digital Sky Survey Data Release 14 (SDSS DR14: \cite{2018A&A...613A..51P}).
We constructed a cubic complete subsample of quasars with a volume of 1 [$\mbox{Gpc}^3$] and performed the PH.
We discovered a characteristic hole (a hollow shell) at a scale $r \sim 150 \; [\mbox{Mpc}]$. 
This exactly corresponds to the BAO signature imprinted in the galaxy/quasar distribution. 
We performed this analysis on a small subsample of 2000 quasars.
This clearly demonstrates that the PH analysis is very efficient in finding this type of topological structures even if the sampling is very sparse. 
}

\begin{document}
\maketitle
\flushbottom

\section{Introduction}
\label{sec:intro}

\subsection{Characterizing the Matter Distribution in the Universe}

The distribution of matter and galaxies at each epoch of the Universe contains fundamental information on the formation and evolution of the cosmic structures in general \citep[e.g.,][]{1980lssu.book.....P, 
1983FCPh....9....1E,2002PhR...367....1B}.   
A vast number of sophisticated methods have been proposed to characterize the statistical properties of the fluctuation in the Universe \citep[e.g.,][]{1980lssu.book.....P,martinez2001statistics,
2002PhR...367....1B}. 
Among them, the $n$-point correlation functions are the most popular and well-studied methods in the analysis of observational galaxy distribution \citep[e.g.,][]{1980lssu.book.....P}. 
Mathematically, the infinite sequence of $n$-point correlation functions are sufficient to specify stochastic random fields \citep[e.g.,][]{adler1981geometry,adler2009random}. 

However, in practice, estimating higher-order $n$-point correlations becomes unrealistically difficult, and other approaches that can treat the information of higher-order correlations have been considered.
One of the popular methods in cosmology and astrophysics is the so-called "Minkowski Funcionals" \citep[e.g.,][]{2003PASJ...55..911H,2003ApJ...584....1M,2019MNRAS.485.1708S}. 
In the applications of the Minkowski Functionals to the Large-Scale Structure, the topology of smoothed density field of galaxy distribution is analyzed by examining the "genus" or number of holes in the density field. 
Since basically the smoothed random field contains the whole information on the infinitely high-order moments of the field, in principle the Minkowski Functionals can specify the galaxy density field completely.
Indeed this method has been applied very extensively to the analysis of the cosmic density field \citep[e.g.,][]{1994A&A...288..697M, 1997MNRAS.284...73K,2002LNP...600..238B,2003PASJ...55..911H,2010A&A...509A..57K,2014MNRAS.443..241W,2017MNRAS.467.3361W}, as well as redshift distortion \citep[e.g.,][]{2018ApJ...863..200A}, weak lensing cosmology \citep[e.g.,][]{2018PhRvD..98j3507S,2019PhRvD..99d3534V,2020A&A...633A..71P}, cosmic reionization \citep[][]{2017MNRAS.465..394Y,2018MNRAS.477.1984B,2019MNRAS.485.2235B,2019ApJ...885...23C}, test of cosmological/gravity theories \citep[][]{2015PhRvD..92f4024L,2015PhRvD..92d3505J,2016PhRvD..94d3506S,2017PhRvL.118r1301F}, and the cosmic microwave background (CMB: e.g., \cite{2010PhRvD..81h3505M,2016A&A...594A..17P,2016JCAP...07..029S,2017PhLB..771...67C,2017JCAP...06..023G,2019JCAP...01..009J}). 

\subsection{From Minkowski Functionals to topological data analysis}

In fact, however, the method of Minkowski Functionals, or sometimes called the genus statistics, is a part of more general framework to characterize the properties of a point data set, known in the field of topological data analysis (TDA: e.g., \cite{edelsbrunner2002,10.1145/997817.997870,wasserman2018}). 
We briefly overview the method of TDA in the following. 

Intuitively, the topological information is characterized by the holes of the object. 
For the current interest, the object to be analyzed is constructed from $n$-dimensional sphere with radius $r$ from discrete data points, i.e., $N$-body particles and/or galaxies (and some other astronomical objects, in general).
With the discovery of the 3-dim structure in galaxy distribution \citep[e.g.,][]{1986ApJ...302L...1D}, an idea of using the geometry (topology) of smoothed density field was proposed to quantify the large scale structure and specify the physics behind it \citep[e.g.,][]{1983ComAp..10...33S,1986ApJ...306..341G,1987ApJ...319....1G,1989ApJ...340..625G,1989A&A...224....1B}.  
The basic idea of these early works was an application of random percolation: they set a sphere around each galaxy with an increasing radius, and examine the geometry of connected components in a considered volume. 
This approach has been developed in relation to the theoretical development of perturbation theory of the cosmic structure \citep[e.g.,][]{1994ApJ...434L..43M,1996ApJ...457...13M,1996ApJ...460...51M,1996ApJ...463..409M,2003ApJ...584....1M,2009PhRvD..80h1301P,2012PhRvD..85b3011G}. 
The expectation of the number of genus of the smoothed field as a function of the threshold density is referred to as the genus curve, and it can be expressed analytically with these theories.
Since this curve can distinguish between Gaussian and other random fields, it became one of the most popular tools for the analysis of the large-scale structure.

\begin{figure}
    \centering
    \includegraphics[width=0.9\textwidth]{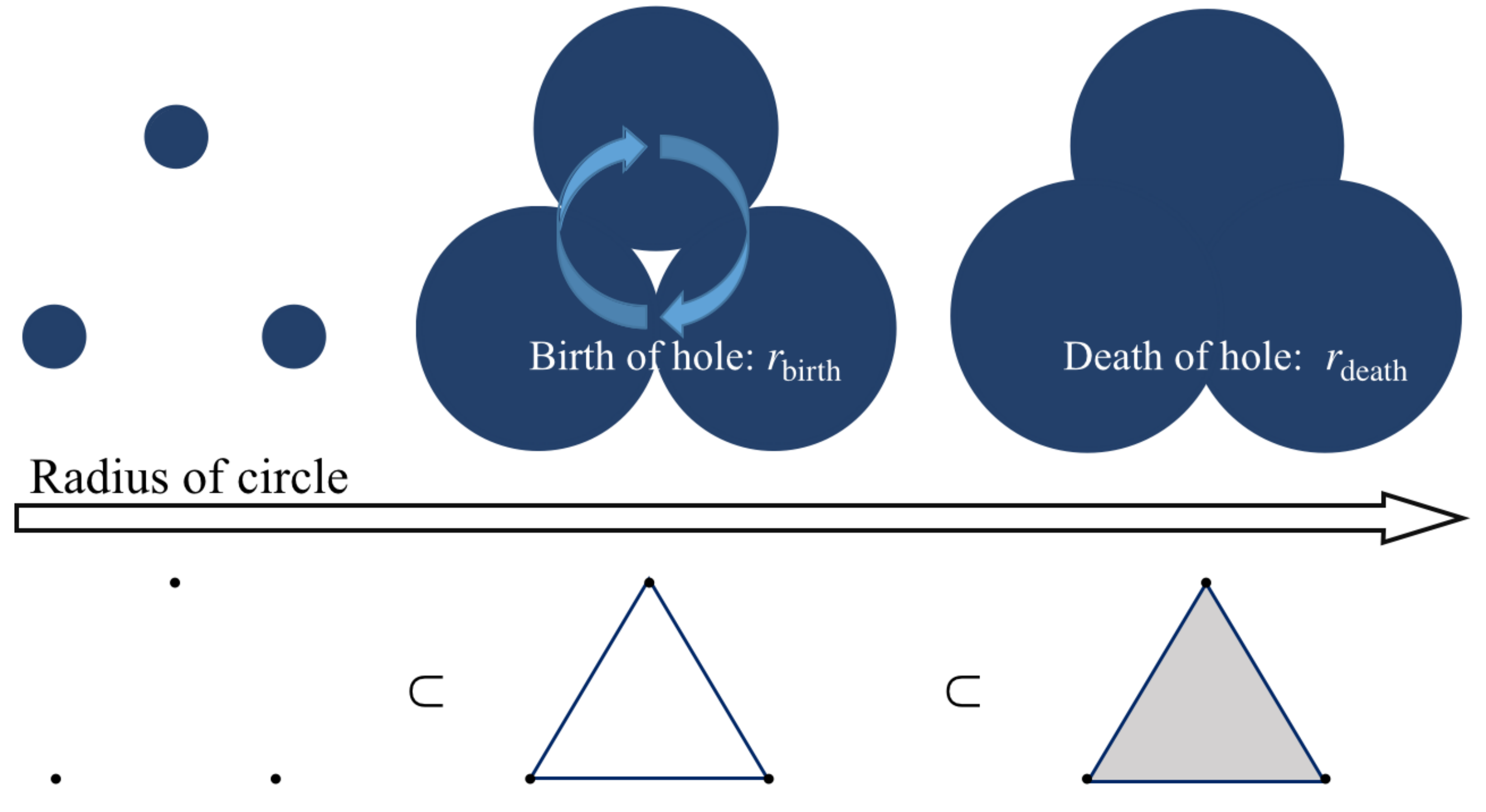}
    \caption{A schematic description on the concept of the persistent homology. 
    Topological information is characterized by "holes" constructed from $n$-dimensional spheres with a radius $r$ from discrete data points.
}
    \label{fig:persistent_homology_schematic}
\end{figure}

In spite of the great development of the theories of the Minkowski Functionals in cosmology, it was not well recognized that it consists a part of much larger, richer framework studied in the topological data analysis (TDA). 
The birth of the primitive concept of the TDA dates back to 90's, but its actual birth was in the new millennium \citep[see e.g., ][]{edelsbrunner2002,DBLP:books/daglib/0025666}.  
The concept directly related to the Minkowski Functionals is the persistent homology (PH) in TDA. 
Similarly to the random percolation, the PH considers a sphere (solid ball) of radius $r$ around each point in a point data set. 
More sophisticatedly, the PH handles the birth and death of loop structures in various dimensions. 
This is schematically described in Fig.~\ref{fig:persistent_homology_schematic}. 
Consider a set of three points.
When the radius $r$ is small, we simply have a set of three balls (left of Fig.~\ref{fig:persistent_homology_schematic}.
Then, at some $r$, we have a connected structure, a loop (middle), regarded as the birth of a hole. 
When $r$ becomes larger, the hole in the loop disappears, the death of the hole (right). 
For general point data set, we trace such a life history of loops with dimension $0, 1, 2, 3, \dots, q$, corresponding to a number of connected components (0-dim), number of loops (1-dim), number of hollow voids (2-dim), $\dots$, number of $q$-dim holes, respectively.  
Thus, we can easily imagine that, if we plot the birth and death of these quantities as a function of $r$, this will contain very rich information of the point set.
Such a plot is referred to as the persistent diagram (PD) and is the central tool for the PH. 
Complete definitions and methodology will be thoroughly presented in Section~\ref{sec:TDA}. 

In spite of the vast number of applications of Minkowski Functionals as mentioned above, extensive applications of the TDA has not been very frequently found in astrophysics and cosmology yet, though it has been gradually known for recent several years. 
Pranav et al.~\cite{2017MNRAS.465.4281P,2019MNRAS.485.4167P} and Feldgrugge et al.~\cite{2019JCAP...09..052F} presented a detailed formulation on the stochastic homology of Gaussian and non-Gaussian random fields, gearing at the application to the stochastic field. 
By the method, they claimed a detection of an anomalous behavior the the CMB fluctuation \citep{2019A&A...627A.163P}. 
Cole and Shiu~\cite{2018JCAP...03..025C} also developed a method to apply the TDA to the evaluation of the non-Gaussianity of the CMB. 
Xu et al.~\cite{2019A&C....27...34X} applied the PH for a large galaxy data set generated from $N$-body simulation \citep{2018MNRAS.473.1195L} and found 23 voids in the simulation data.

Here, in this work, we introduce the method with the emphasis on the {\it inverse analysis} of the detected topological structures, and apply the PH to the signal of the baryon acoustic oscillation. 

\subsection{The baryon acoustic oscillation}

Baryons evolve in a very complicated way via electromagnetic interactions (with radiative heating/cooling, gas pressure, fluid dynamical processes, etc.)
A typical example of such a nontrivial phenomenon is the baryon acoustic oscillation (BAO) generated in the matter-radiation fluid in the early Universe \citep{1970ApJ...162..815P,1970Ap&SS...7....3S}. 
Here we briefly review the BAO and related observables. 

Consider a point-like initial perturbation in the primordial matter-photon plasma. 
In the plasma, the matter and photons are locked into a single fluid. 
Since the photons are so hot and numerous, the combined fluid has an tremendous pressure with respect to its density. 
The pressure tries to equalize itself with the surroundings, which results in an expanding spherical sound wave. 
The sound speed at this early epoch is evaluated by 
\begin{eqnarray}\label{eq:sound_speed_decoupling}
  c_{\rm s} = \frac{\displaystyle c}{\displaystyle \sqrt{3\left( 1 + \frac{3\rho_{\rm B}}{4\rho_{\rm R}} \right)}}
\end{eqnarray}
where $c$ is the speed of light, and the subscript B and R stand for baryon and radiation, respectively. 
Equation~\ref{eq:sound_speed_decoupling} shows that the sound speed  before the matter-radiation decoupling can be  $\sim 57$~\% of the speed of light at early epoch.  
Sound horizon (final radius) $r_{\rm s}$ is obtained by
\begin{eqnarray}
  r_{\rm s} = \int_0^{t_{\rm dec}} (1+z) c_{\rm s} {\rm d} t = 
  \int_{z_{\rm dec}}^{\infty} \frac{c_{\rm s} }{H_0\sqrt{\Omega_{\rm R,0}(1+z)^4+\Omega_{\rm M,0}(1+z)^3 + \Omega_{\Lambda, 0}}}{\rm d} z
\end{eqnarray}
where subscript dec stands for the time of the decoupling between photons and matter. 
What we actually observe is the superposition of many acoustic waves imprinted on the large-scale structure emerged from the primordial fluctuations.
Though the BAO is a baryonic phenomenon by definition, the density enhancement structure of baryons and dark matter becomes the same at the final state \citep[e.g.,][]{MaBertschinger:1995, 2007ApJ...664..675E}.

The BAO length scale is constant in comoving coordinates. 
Then, in principle, we can detect the signal on the galaxy 2-point correlation function. 
However, since the BAO scale is very large compared to the typical scale of the large-scale structure, we need a very large galaxy sample with dense sampling, since we must measure the signal at such a large scale, where usually the 2-point correlation is very weak. 
The SDSS is the largest optical photometric and spectroscopic surveys ever existed, covering one-third of the whole sky. 
Then, the advent of the SDSS finally made it realistic to detect the BAO signal on the 2-point correlation function. 
Eisenstein et al.~\cite{2005ApJ...633..560E} first detected the signal around 150 Mpc on the 2-point correlation function. 
Since then, still this analysis is only possible with SDSS data. 

The BAO can be used for various directions of cosmological studies. 
Then, if a more flexible and easier method for the detection and quantifying the BAO signal exists, it would be desirable. 
We introduce the PH as such an ideal method for the analysis of BAO. 
This paper is organized as follows: 
We introduce the definitions and concepts of the TDA relatively rigorously in Section~\ref{sec:TDA}. 
Especially, we introduce the method of inverse analysis of the PH. 
Then we examine the performance of the PH as a tool for the studies of BAO by using simulation datasets with and without baryon physics in Section~\ref{sec:simulation}. 
After knowing its power, we apply the PH to quasar data extracted from the extended Baryon Oscillation Spectroscopic Survey (eBOSS) \cite{2016AJ....151...44D} LSS catalog \cite{2018MNRAS.473.4773A} in the SDSS DR14 \cite{2018A&A...613A..51P} in Section~\ref{sec:SDSS}. 
Section~\ref{sec:summary} is devoted to our summary and discussion on the result and future prospects of the presented methodology. 

In this study, we adopt cosmological parameters $h = H_0/100\ [{\rm km\ s^{-1}\ Mpc^{-1}}]= 0.6766$, $\Omega_{\Lambda,0} = 0.6889$, $\Omega_{\rm M,0}=0.3111$, and $\ln (10^{10} A_s) = 3.047$, as constrained by the latest CMB observation by Planck \cite{Planck+CosPar:2018} throughout this paper.

\section{Topological Data Analysis (TDA)}\label{sec:TDA}

\subsection{Geometric modeling of point cloud data}

In the topological data analysis (TDA), we often face a problem to examine topological properties of a finite sequence of points $P$ in $N$-dim Euclidean space $\mathbb{R}^N$\footnote{Throughout this paper, $\mathbb{R}, \mathbb{N}$, and $\mathbb{N}_0$ stand for a set of real number, set of natural number, and non-negative natural number, respectively.}.  
Since $P$ is a finite set, a simple application of traditional topological discussion cannot extract any nontrivial information on the features of $P$. 
Instead, in TDA, a geometric model with a "filtration", $\mathcal{X} = \{ X^a ; a \in \mathbb{R} \}$, is constructed from the finite point cloud, and  topological methods are applied to the model. 
Here the parameter of the filtration $a$ is considered to control the "resolution" or "scale" of the input data.
The TDA extract meaningful topological characteristics of the point data by examining the persistent properties along the change of the resolution \citep[e.g.,][]{edelsbrunner2002,boissonnat_chazal_yvinec_2018}. 
We introduce such geometric models that play a fundamental role in the TDA. 

\subsubsection{Simplex and simplicial complex}

\begin{figure}
    \centering
    \includegraphics[width=0.8\textwidth]{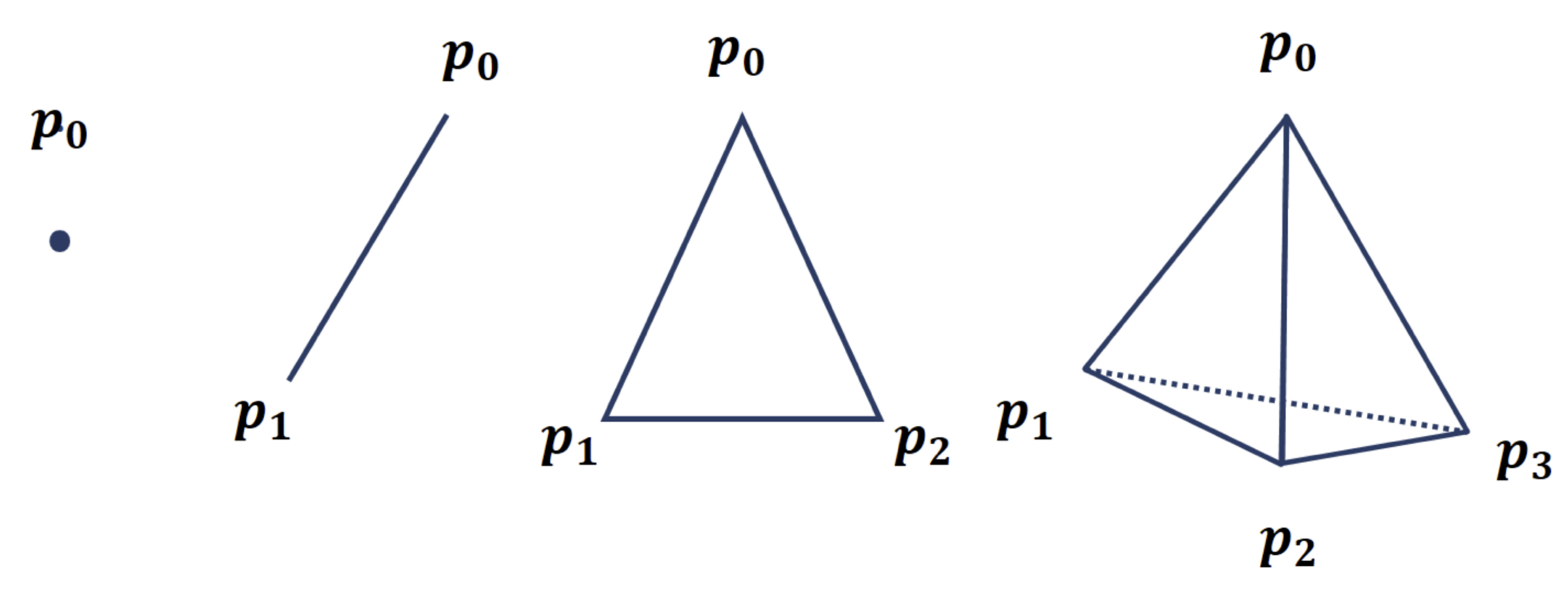}
    \caption{An example of 0, 1, 2, and 3-dim simplex. }
    \label{fig:simplex}
\end{figure}

A simplex is a generalization of the notion of a triangle or tetrahedron to arbitrary dimensions.
More precisely, it is defined as follows. 
Points of a finite set $P = \{ p_0, p_1, \dots, p_k \}$ in $\mathbb{R}^N$ is affinely independent if they are not contained in any affine subspace of dimension less than $k$.   
\begin{definition}[Simplex]~\\
Suppose a set of $k+1$ points, $P = \{ p_i\}\; (i = 0, \dots, k)$ in $N$-dimensional Euclidean space $\mathbb{R}^N$, arranged to be affinely independent. 
If we consider $k$ vectors $\overrightarrow{p_0 p_1}, \dots, \overrightarrow{p_0 p_k}$, the convex hull that contains these vectors
\begin{eqnarray}
  \left| p_0 p_1\dots p_k \right| \equiv \left\{ x \in \mathbb{R}^N | x = \lambda_0 p_0 + \dots + \lambda_k p_k, \lambda_i \geq 0\,, \;  \sum_{i=0}^{k} \lambda_i = 1 \right\}
\end{eqnarray}
is called $k$-simplex, often denoted as $\sigma$. 
The dimension of $k$-simplex is defined as $k$. 
\end{definition}
For example, a $0$-simplex is a point, a $1$-simplex is a line segment, a $2$-simplex is a triangle, and a $3$-simplex is a tetrahedron. 

\begin{definition}[Face]~\\
Consider a $k$-simplex. 
The convex hull of any nonempty subset of the $\ell+1$ points ($\ell \leq k$) that define an $\ell$-simplex is called a face of the simplex.
\end{definition}
A face is itself a simplex. 
The convex hull of a subset of size $\ell+1$ of the $k+1$ points is defined as $\ell$-simplex, called an $\ell$-face of the $k$-simplex. 
A $0$-face is called the vertex, and 
a 1-face is the edge. 
Figure~\ref{fig:simplex} shows the examples of simplices with dimension of 0, 1, 2, and 3. 
For example, a $2$-simplex in Fig.~\ref{fig:simplex} has seven surfaces
$|p_0|, |p_1|, |p_2|, |p_0 p_1|, |p_0 p_2|, |p_1 p_2|$, and $|p_0 p_1 p_2|$. 

\begin{figure}
    \centering
    \includegraphics[width=0.4\textwidth]{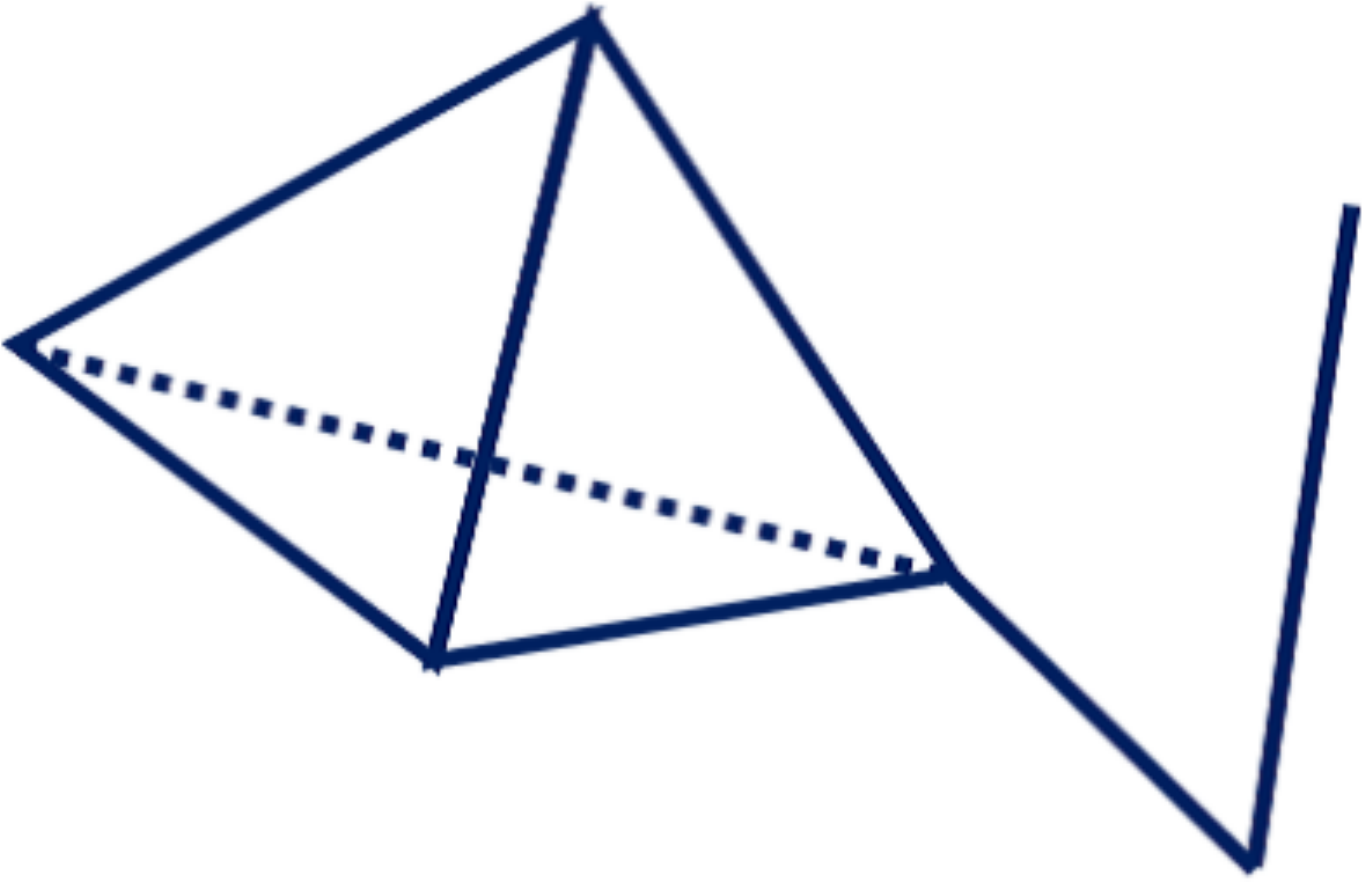}
    \caption{An example of simplicial complex. }
    \label{fig:simplicial_complex}
\end{figure}

Based on these concepts, we can define the simplicial complex as
\begin{definition}[Simplicial complex]\label{def:simplicial_complex}
~\\
A simplicial complex $K$ in $\mathbb{R}^N$ is a set of simplices that satisfies the following conditions:
\begin{enumerate}
    \item Every face $\sigma$ of a simplex from $K$ is also in $K$.
    \item The non-empty intersection of any two simplices $\sigma_{i}, \sigma_{j} \in K$ is a face of both $\sigma _{i}$ and $\sigma_{j}$ $(0 \leq i, j \leq k)$.
\end{enumerate}
\end{definition}
The maximal dimension of simplices in $K$ is called the dimension of the simplicial complex $K$, denoted as $\mbox{dim}\;K$. 
An example of a simplicial complex is presented in Fig.~\ref{fig:simplicial_complex}. 

\begin{definition}[Polyhedron]~\\
Consider a union of all the simplices in $K$, 
\begin{eqnarray}
  |K| \equiv \bigcup_{\sigma \in K} \sigma \, , 
\end{eqnarray}
a geometrical shape in $\mathbb{R}^N$ is obtained. 
This shape is referred to as a polyhedron defined by a simplicial complex $K$. 
\end{definition}

\begin{definition}[Abstract simplicial complex]~\\
If a family $\tilde{K}$ of non-empty finite subsets of a finite set $V$ satisfies the following condition, $(V, \tilde{K})$ is an abstract simplicial complex:
\begin{enumerate}
    \item For every set $v \in V$, $\{ v\} \in \tilde{K}$, 
    \item A subset of $\tau \subset \tilde{K}$, $\sigma \subset \tau$, also belongs to $\tilde{K}$.
\end{enumerate}
\end{definition}
Here a subset $\tau$ of $\tilde{K}$, $\tau = \{ v_0, \dots, v_k\}$, is called $k$-simplex, whose dimension is defined as $k$. 
The maximal dimension of simplices in $\tilde{K}$ is called the dimension of the abstract simplicial complex, denoted as $\mbox{dim}\; (V, \tilde{K})$. 
As trivially follows from the definition, a simplex consisting an abstract simplicial complex does not need to be in $\mathbb{R}^N$. 

A simplicial complex can be treated as an abstract simplicial complex. 
Inversely, a simplicial complex in $\mathbb{R}^N$ can be assigned to an abstract simplicial complex. 
Consider a point
\begin{eqnarray}
  p_i = (0, \dots, 0, \underset{i}{1}, 0, \dots, 0) \in \mathbb{R}^N
\end{eqnarray}
(i.e., only the $i$-th component is $1$ and all the others are $0$) for each element $v_i$ ($i = 1, \dots, N$) of an abstract simplicial complex $(V, \tilde{K})$.
By these points $\{ p_i\}$, we define a simplex $|p_0, \dots, p_k|$ for  $\{v_0, \dots, v_k \} \in \tilde{K}$. 
If we define $K$ as a set of all the simplices in $\mathbb{R}^N$ obtained as above, $K$ satisfies Definition~\ref{def:simplicial_complex}.
Namely, $K$ is a simplicial complex. 
The polyhedron $|K|$ is called a geometric realization of $(V, \tilde{K})$ (e.g., \cite{boissonnat_chazal_yvinec_2018}).  

\subsubsection{Nerve and nerve theorem}\label{subsec:nerve}

As mentioned above, simplicial complices can be seen as purely combinatorial objects, as well as a topological space at the same time. 
We introduce an important concept "nerve" here. 
\begin{definition}[Open cover]\label{def:open_cover}~\\
An open cover of a set $X$ is a collection $\mathcal{U} = \{ U_i \}\; (i = 1, \dots, m)$ of open subsets $U_i \subseteq X$, such that
\begin{eqnarray}
  X = \bigcup_{i=1}^{m} U_i \; .
\end{eqnarray}
\end{definition}
Then, we define a nerve as follows. 
\begin{definition}[Nerve]\label{def:nerve}~\\
Given a cover of a set $X$, $\mathcal{U} = \{ U_i \} \in \mathbb{R}^N \; (i=1, \dots, m)$, its nerve is the abstract simplicial complex $\tilde{K} (\mathcal{U})$ whose vertex set is $\mathcal{U}$ and 
\begin{eqnarray}
  \sigma = [ U_{i_0}, U_{i_1}, \dots, U_{i_k}] \in \tilde{K}(\mathcal{U}) 
\end{eqnarray}
if and only if 
\begin{eqnarray}
  \bigcap_{j=1}^{k} U_{i_j} \neq \emptyset \; .
\end{eqnarray}
We denote the nerve of $\mathcal{U}$ as $\mathcal{N}(\mathcal{U})$. 
\end{definition}
In general, the nerve $\mathcal{N}(\mathcal{U})$ does not always accurately reflect the topology of $X$, i.e., $\mathcal{N}(\mathcal{U})$  is not homotopic equivalent to $X$.
However, if an intersection of the $U_i$s is either empty or contractible (see Definition~\ref{def:contractible}), namely a "good cover", the following important theorem holds.
\begin{theorem}[The Nerve Theorem]\label{theo:nerve}~\\
Let $\mathcal{U} = \{ U_i \} \in \mathbb{R}^N \; (i=1, \dots, m)$ be a finite open cover of a subset $X \in \mathbb{R}^N$ such that any intersection of the $U_i$s is either empty or contractible. 
Then $X$ and the nerve $\mathcal{N}(\mathcal{U})$ are homotopy equivalent.
\end{theorem}

\subsubsection{\v{C}ech complex}\label{subsubsec:cech_complex}
\begin{figure}
    \centering
    \includegraphics[width=0.3\textwidth]{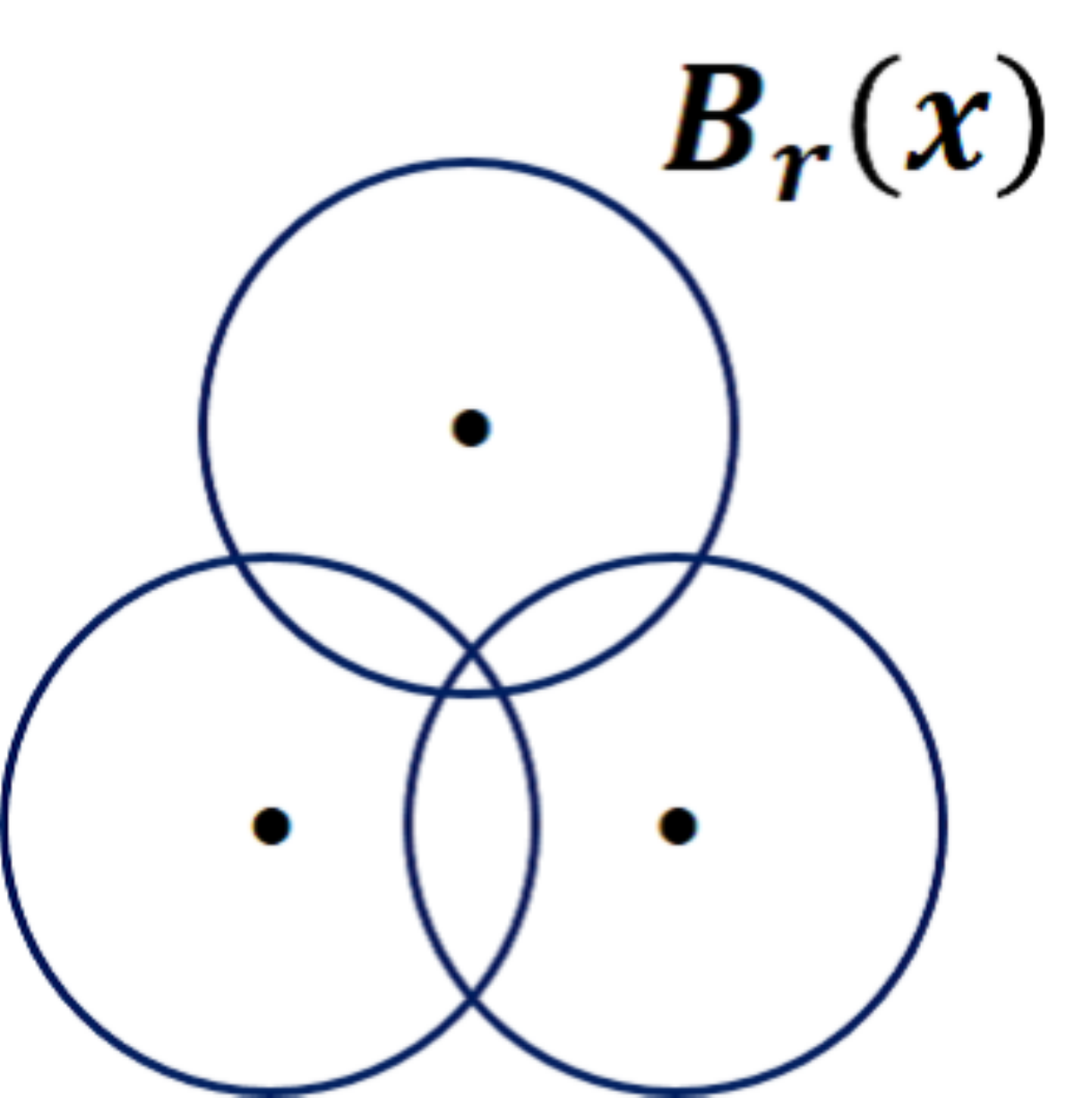}
    \includegraphics[width=0.3\textwidth]{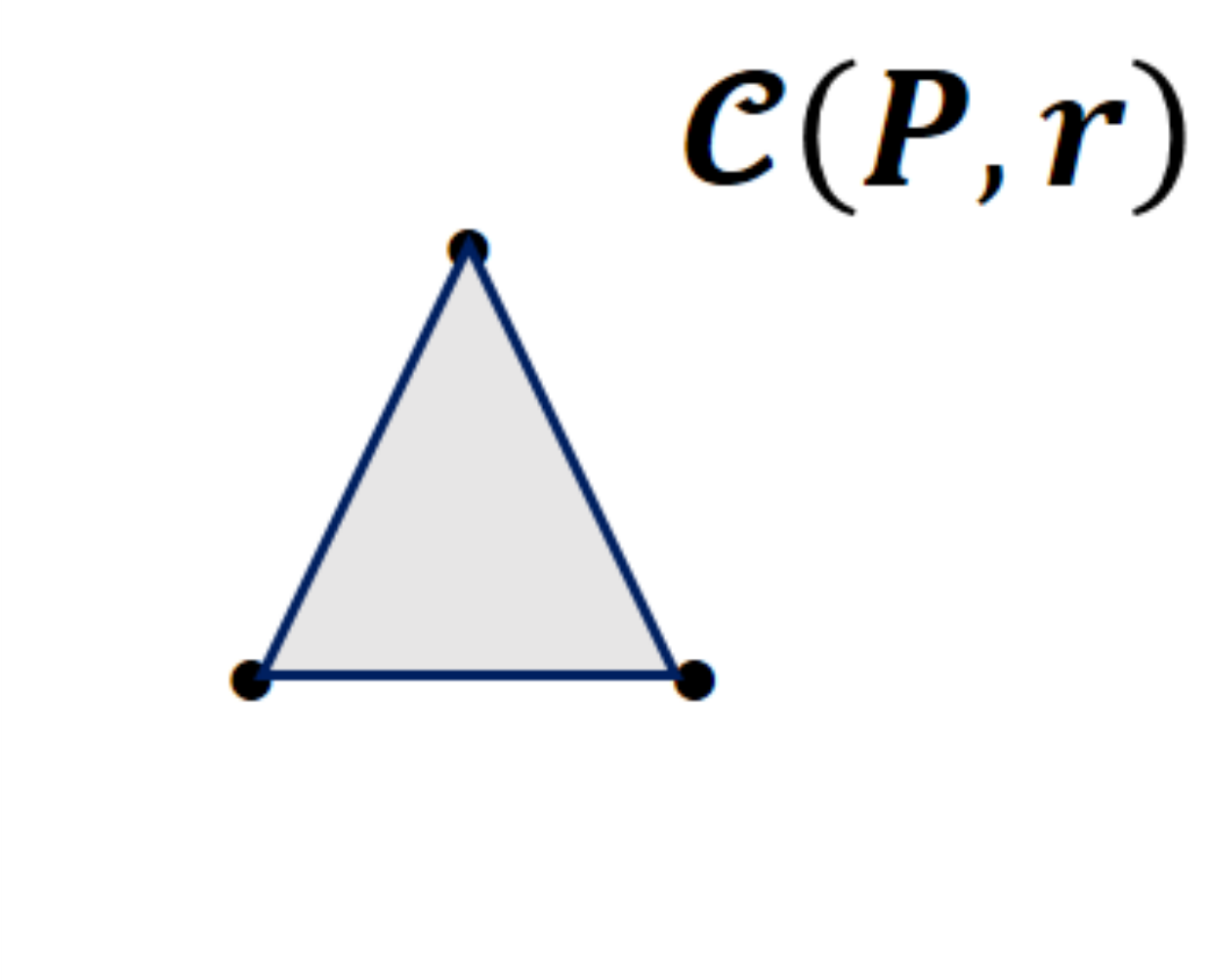}
    \caption{An example of \v{C}ech complex. 
    Left: a union of balls with radius $r$ around $x_i$, Right: the corresponding \v{C}ech complex. }
    \label{fig:cech_complex}
\end{figure}

Now we relate the filtration and simplicial complex. 
\begin{definition}[Filtration of simplical complices]~\\
A filtration of a finite simplicial complex $K$ is a nested sequence of sub-complices $\emptyset \subset K^1 \subset \dots \subset K^m = K$ such that
\begin{eqnarray}
  K^{i+1} = K^i \cup \sigma^{i+1} 
\end{eqnarray}
where $\sigma^{i+1}$ is a simplex of $K$. 
\end{definition}
We note that a filtration of $K$ is just an ordering of the simplices, it may be natural to index the simplices by an increasing sequence of real numbers $\{ \alpha_i \} \in \mathbb{R} \, , \; (i = 1, \dots, m)$,
\begin{eqnarray}
  \emptyset \subset K^{\alpha_0} \subset K^{\alpha_1} \subset \dots \subset K^{\alpha_m} = K. 
\end{eqnarray}

We define the \v{C}ech complex as follows. 
\begin{definition}[\v{C}ech complex]\label{def:cech_complex}~\\
Consider a point set $P \in \mathbb{R}^N$, $P = \{ x_i \in \mathbb{R}^N, i = 1, \dots, m\}$. 
Set a sphere with radius $r$ around each $x_i$, 
\begin{eqnarray}
  B_r (x_i) = \left\{ x \in \mathbb{R}^N;  \|x - x_i\| \leq r \right\} \; ,
\end{eqnarray}
where $\|x\|$ stands for the Euclid norm. 
The nerve of the collection of these balls $\mathcal{B} = \{ B_r(x_i), x_i \in X \}$, 
\begin{eqnarray}
  \mathcal{C}(P, r) \equiv \mathcal{N}(\mathcal{B}) = \left\{ |x_{i_0}, \dots, x_{i_k}|; \bigcap_{j=0}^{m} B_r(x_{i_j}) \neq \emptyset \right\} 
\end{eqnarray}
is the \v{C}ech complex of $P$ with radius $r$. 
\end{definition}
Since a ball satisfies the condition of Theorem~\ref{theo:nerve}, we have
\begin{eqnarray}
  X_r \equiv \bigcup_{i=1}^{m} B_r(x_i) \simeq \mathcal{C}(P,r) \; .
\end{eqnarray}
This means that if we want to examine homotopy invariant characteristics of a set $X$ that satisfies the condition of Theorem~\ref{theo:nerve}, we can do it equivalently on its nerve $\mathcal{N}(\mathcal{U})$. 
Since the nerve is an abstract simplicial complex, it is much more computer-friendly to calculate such properties. 
A schematic example of a \v{C}ech complex is shown in Fig.~\ref{fig:cech_complex}

The necessary and sufficient condition that $k$-simplex of a \v{C}ech complex is 
\begin{eqnarray}
  \bigcap_{j=0}^{k} B_r(x_{i_j}) \neq \emptyset \; .
\end{eqnarray}
Note that this hols for $r' > r$, namely, 
\begin{eqnarray}
  \bigcap_{j=0}^{k} B_r(x_{i_j}) \neq \emptyset \; \Rightarrow 
  \bigcap_{j=0}^{k} B_{r'}(x_{i_j}) \neq \emptyset , \quad (r < r') \; .
\end{eqnarray}
Thus
\begin{eqnarray}
  \mathcal{C} (P, r) \subset  \mathcal{C} (P, r')  , \quad (r < r') \; ,
\end{eqnarray}
and therefore, for an increasing sequence $r_0 < r_1 < \dots < r_T$, we have an increasing sequence of \v{C}ech complices
\begin{eqnarray}
  \mathcal{C} (P, r_0) \subset \dots \subset \mathcal{C} (P, r_i) \subset \dots \subset \mathcal{C} (P, r_T) \; .
\end{eqnarray}
This defines the \v{C}ech complex filtration.
This enables us to deal not only with the topological information of $\mathcal{C}(P, r_i)$ at a radius $r_i$ but also their transition and persistence with changing $r$. 

\subsubsection{Voronoi diagrams, Delauney complex and alpha complex}

\begin{figure}
    \centering
    \includegraphics[width=0.9\textwidth,angle=0]{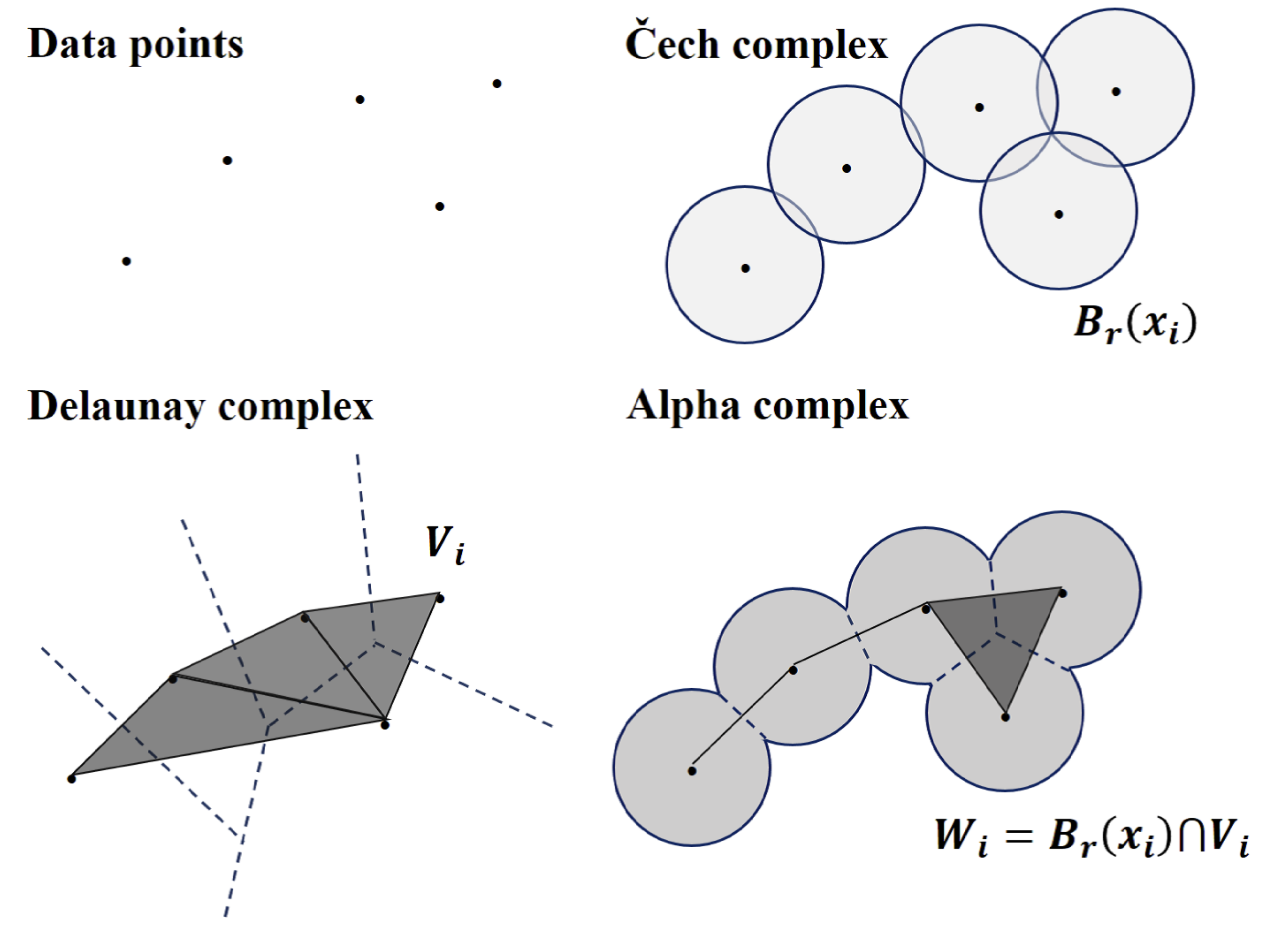}
    \caption{Schematic description of the relation between a point set (top left), its \v{C}ech complex (top right), Delauney complex (bottom left), and alpha complex (bottom right). }
    \label{fig:summary_complex}
\end{figure}

Though the definition of \v{C}ech complex is straightforward, computationally it is rather demanding when the data size is large. 
A more convenient type of complex is desirable for this aspect. 
We introduce the alpha complex in the following \cite{10.1145/174462.156635,10.1145/160985.161139}. 
We first define the Voronoi diagram of a point set. 

\begin{definition}[Voronoi diagram]\label{def:voronoi}~\\
Consider a point set $P \in \mathbb{R}^N$, $P = \{ x_i \in \mathbb{R}^N, i = 1, \dots, m\}$. 
A Voronoi diagram can be expressed with Voronoi cells $V_i\; (i=1, \dots, m)$ for each point $x_i$ as
\begin{eqnarray}
  &&V_i = \left\{ x \in \mathbb{R}^N;  \|x - x_i\| \leq \|x - x_j\| , 1 \leq j \leq m, j \neq i \right\} \; , \\
  &&\mathbb{R}^N = \bigcup_{i=1}^{m} V_i \; . \label{eq:voronoi_tessellation}
\end{eqnarray}
A region defined for each $x_i$ is the Voronoi cell, and the tessellation (eq.~(\ref{eq:voronoi_tessellation})) is referred to as the Voronoi diagram. 
\end{definition}

Then, we define Delauney complex. 
\begin{definition}
Delauney complex $\mathcal{D}(P)$ is the nerve $\mathcal{N}(\mathcal{V})$ of the Voronoi diagram,
\begin{eqnarray}
  \mathcal{D} (P) = \mathcal{N}(\mathcal{V})
\end{eqnarray}
where $\mathcal{V}$ is a convex closed set
\begin{eqnarray}
  \mathcal{V} = \{ V_i ; i = 1, \dots, m\} \; .
\end{eqnarray}
\end{definition}

Further, we consider the most important concept for this analysis.
\begin{definition}[Alpha complex]~\\
We define an intersection of $B_r(x_i)$ and $V_i$ denoted as $W_i$,
\begin{eqnarray}
  &&W_i \equiv  B_r(x_i) \cap V_i \; , \\
  &&X_r = \bigcup_{i=1}^{m} W_i \; .
\end{eqnarray}
The alpha complex $\mathcal{A}(P, r)$ for a set of the center of spheres $P$ can be defined as
\begin{eqnarray}
  \mathcal{A} (P, r) \equiv \mathcal{N} (\mathcal{W}) \; , 
\end{eqnarray}
where $\mathcal{W} = \{ W_i ; i=1, \dots, m\}$.
\end{definition}
The motivation to introduce the Alpha complex was to introduce a simplicial complex model to connect a point sequence and its convex hull with spatial resolution \cite{10.1145/174462.156635,10.1145/160985.161139}. 
We note that the dimension of $\mathcal{A}(P, r)$ is at most the dimension $N$ of the embedding space.  

\subsection{Simplicial homology}
We introduce the basic notions of simplical homology that are necessary to define topological persistence. 
According to \cite{boissonnat_chazal_yvinec_2018}, we restrict the homology with coefficients in the finite field $\mathbb{Z}_2 \equiv \mathbb{Z}/2\mathbb{Z}=\{0, 1 \}$. 
In this work, we did out best to avoid to introduce complicated algebraic concepts. 
Readers who prefer a rigorous introduction are guided to read some textbooks on abstract algebra \citep[e.g.,][]{hatcher2002,dummit2004}. 

\subsubsection{Space of $k$-chains}

For any non-negative integer $k \in \mathcal{N}_0$, the space of $k$-chains is a vector space of all the formal sums with coefficients in $\mathbb{Z}_2$ of $k$-dim simplices of $K$. 
Precisely, if $\{ \sigma_1, \dots, \sigma_p\}$ is a set of $k$-simplices of $K$, any $k$-chain is uniquely written as
\begin{eqnarray}
  c = \sum_{i=1}^{p} \varepsilon_i \sigma_i 
\end{eqnarray}
with $\varepsilon_i \in \mathbb{Z}_2$. 
\begin{definition}[Space of $k$-chains]~\\
The space of $k$-chain is a set of $C_k(K)$ of the simplicial $k$-chains of $K$ with operations of a sum and a scalar product as
\begin{eqnarray}
c+c' &=& \sum_{i=1}^{p} (\varepsilon_i +\varepsilon_i') \sigma_i\; , \\
\lambda  c &=& \sum_{i=1}^{p} (\lambda \varepsilon_i) \sigma_i  \, .
\end{eqnarray}
This forms a $\mathbb{Z}_2$-vector space whose zero element is 
\begin{eqnarray}
  0 = \sum_{i=1}^{p} 0 \sigma_i \; .
\end{eqnarray}
\end{definition}
Note that the sums and products are modulo 2. 
The set of $k$-simplices of $K$ consists the basis of $C_i(K)$.

Chains with coefficient in $\mathbb{Z}_2$ have a straightforward geometric interpretation.
Since any $k$-chain is uniquely written as $c = \sigma_{i_1} + \dots + \sigma_{i_m}$ where $\sigma_{i_j}$ are $k$-simplices, $c$ can be regarded as the union of the simplices $\{ \sigma_{i_j}\}$.  

\subsubsection{Boundary operator and homology group}

\begin{definition}[Boundary of a simplex]~\\
A boundary $\partial(\sigma)$ of a $k$-simplex $\sigma$ is the sum of its $(k-1)$-faces. 
This is a $(k-1)$-chain. 
let $\sigma_i = \left[ v_0, \dots, v_k \right]$ a $k$-simplex, then
\begin{eqnarray}
\partial \sigma_i \equiv \sum_{i=0}^{k} \left[ v_0, \dots, \check{v_i}, \dots, v_k \right]
\end{eqnarray}
($\check{v_i}$ means $i$-th component is removed).
\end{definition}

The boundary operator defined on the simplices of $K$ can be extended linearly to $C_k(K)$. 
\begin{definition}[Boundary operator]\label{def:boundary}~\\
The boundary operator $\partial_k: C_k(K) \rightarrow C_{k-1}(K)$ is defined as
\begin{eqnarray}
  \partial[v_1,v_2,\cdots,v_n] = \sum^n_{i=0} [v_1,v_2,\cdots,\check{v}_i,\cdots,v_n] \;.
\end{eqnarray}
\end{definition}
The boundary operator satisfies the following important property. 
\begin{theorem}\label{theo:boundary}~\\
The boundary of a boundary of a chain is always zero, i.e., 
For all $k \in \mathbb{N}$, 
\begin{eqnarray}
  \partial_{k-1} \circ \partial_k = 0 \; .
\end{eqnarray} 
\end{theorem}
The boundary operator defines a sequence of linear maps between the spaces of chains. 
\begin{definition}[Chain complex]\label{def:chain_complex}~\\
A chain complex associated with a complex $K$ of dimension $N$ is a sequence of linear operators as
\begin{eqnarray}
  \{ 0 \} \xrightarrow{} C_N(K) \xrightarrow{\partial_N} C_{N-1}(K) \xrightarrow{\partial_{N-1}} \dots \xrightarrow{\partial_{k+1}}C_k\xrightarrow{\partial_{k}}C_{k-1}\xrightarrow{\partial_{k-1}} \cdots \xrightarrow{\partial_{2}}C_1\xrightarrow{\partial_{1}}C_0\xrightarrow{\partial_{0}} \{0\} \,. \nonumber \\
\end{eqnarray}
\end{definition}
For $k \in [0, \dots, N]$, We further introduce the following.
\begin{definition}[Cycle and boundary of $k$-cycles]
\begin{eqnarray}
  Z_k(K) &\equiv& {\rm Ker}\{\partial: C_{k}\to C_{k-1}\} = \{c \in C_k(K); \partial_k(c) = 0 \} \; ,\\
  B_k(K) &\equiv& {\rm Im}\{\partial: C_{k+1}\to C_{k}\} = \{c \in C_k(K); \partial_{k+1}(c'), c' \in C_{k+1}(K)\} \; 
\end{eqnarray}
are the $k$-cycles and $k$-boundaries of a $k$-chain, respectively. 
\end{definition}
By definition, $Z_k$ means it does not have a boundary, and the $k$-boundary is the boundary of a $k+1$-chain. 
It follows from Theorem~\ref{theo:boundary} that 
\begin{eqnarray}
  B_k(X) \subset Z_k(X) \subset C_k(K) \; . 
\end{eqnarray}

Thus, the quotient group $Z_k/B_k$ can measure the difference of the two, and enables us to extract the $k$-complices without boundaries but not boundaries of a set of $k+1$-complices. 
\begin{definition}[Homology group]\label{def:homology_group}~\\
A quotient vector space
\begin{eqnarray}
  H_k(K) = Z_k(K)/B_k(K)
\end{eqnarray}
is the $k$-th homology group of a simplicial complex $K$. 
Elements of the vector space $H_k(K)$ is the homology class of $K$.
The dimension $\beta_k(K)$ of $H_k(K)$ is the $k$-th Betti number of $K$. 
\end{definition}
Informally, the homology class represents a hole\footnote{The homology classes with $k=0, 1, 2$ are mainly discussed in the analysis with persistent homology.  
We expect one would not be confused between the $k=0$ homology class and the Hubble parameter $H_0$. }. 

\subsection{Persistent homology}\label{sec:PH}

The concept of persistent homology was introduce by \cite{edelsbrunner2002}. 
Intuitively, the persistent homology aims at pursuing the track of all 
subcomplices of a filtration, and to pair the creation (birth) and destruction (death) of homology classes appearing during the process. 

We define the persistence for a filtration of a simplical complex. 
Let $K$ be a $N$-dim simplicial complex and let
\begin{eqnarray}
  \emptyset = K^0 \subset K^1 \subset \dots \subset K^m = K
\end{eqnarray}
be a filtration of $K$ such that, for any $i = 0, \dots, m-1$, $K^{i+1} \equiv K^i \cup \sigma^{i+1}$ where $\sigma^{i+1}$ is a simplex. 
For any $0 \leq n \leq m$, we denote the set of $k$-chains of $K$ (with coefficients in $\mathbb{Z}_2$) by $C_k^n$. 
The restriction of the boundary operator to $C_k^n$ has its image contained in $C_{k-1}^{n-1}$. 
We denote the sets of $k$-cycles and $k$-boundaries of $K^n$ by $Z_k^n$ and $B_k^n$, respectively. 
\begin{definition}[Persistent homology]~\\
The $k$-th persistent homology group of $K^n$ is 
\begin{eqnarray}
  H_k^n \equiv Z_k^n/B_k^n \, . 
\end{eqnarray}
\end{definition}
With these definitions, we have
\begin{eqnarray}
  &&Z_k^0 \subset Z_k^1 \subset \dots \subset Z_k^n \subset \dots \subset Z_k^m = Z_k(K) \; , \\
  &&B_k^0 \subset B_k^1 \subset \dots \subset B_k^n \subset \dots \subset B_k^m = B_k(K) \; . 
\end{eqnarray}
\begin{definition}[Persistent Betti number]
For $p \in \{ 0, \dots, m \}$and $\ell \in \{ 0, \dots, m^p \}$, the $k$-th persistent Betti number of $K^\ell$ is the dimension of the vector space
\begin{eqnarray}\label{eq:betti}
  H_k^{\ell, p}(K) = Z_k^\ell/\left( B_k^{\ell + p} \cap Z_k^{\ell}\right) \; .
\end{eqnarray}
Here $Z_k^\ell$ and $B_k^{\ell + p}$ stands for a $k$-cycle group on the chain group of $X^\ell$ and a $k$-boundary group of $X^{\ell+p}$, respectively.
\end{definition}
The $k$-th persistent Betti number of $K^\ell$ represents the number of independent homology classes of $k$-cycles in $K^\ell$ that are not boundaries in $K^{\ell+p}$. 
A $k$-cycle in $K^\ell$ generating a nonzero element in $H_k^{\ell+p}$ is a cycle, that has appeared in the filtration before the step $\ell + 1$ and that is still not a boundary at step $\ell+p$. 
Namely, for a sequence of homology determined from the filtration 
\begin{eqnarray}
  H_k(K^0) \to H_k(K^1) \to \dots \to H_k(K^\ell) \to \dots \to H_k(K^{\ell+p}) \to \dots \; , 
\end{eqnarray}
it can be examined whether an element of $H_k(K^\ell)$ still exists in $H_k(K^{\ell+p})$ that is $p$-steps ahead along the filtration.

Thus, an element of $H_k(X^\ell)$ that survives long is regarded as an important topological feature, while an element that lives short (i.e., during a small interval $\ell$) is treated as topological noise. 
Intuitively, persistent homology is a method for computing topological features of a space at different spatial resolutions. 
More persistent features are detected over a wide range of spatial scales and are more likely to represent "true" features of the underlying point cloud rather than artifacts of sampling, noise, or some particular choice of parameters.

\begin{figure}
    \centering
    \includegraphics[width=0.4\textwidth]{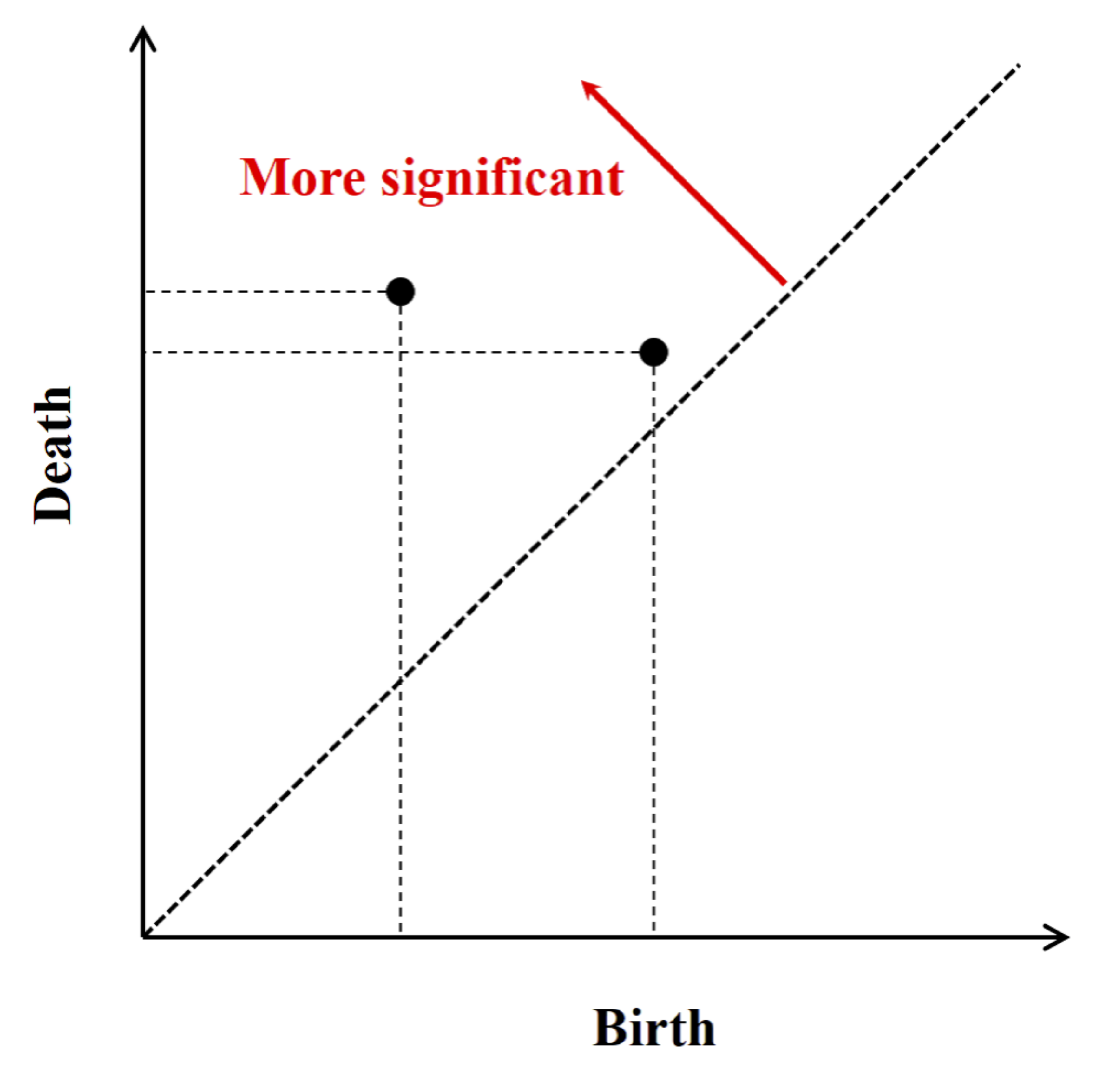}
    \caption{Schematic persistence diagram. 
    Data points laying close to the diagonal line should be regarded as a topological noise, while those far from the diagonal represent significant and robust structures in the data point cloud. }
    \label{fig:PD_schematic}
\end{figure}

\subsection{Persistence diagram}\label{subsec:PD}

\subsubsection{Persistent pair}

To make use of the persistent homology, we define a few more notions.
\begin{definition}[Positive and negative simplex]~\\
Let $K$ be a $d$-sim simplical complex and 
\begin{eqnarray*}
  \emptyset = K^0 \subset K^1 \subset \dots \subset K^m = K \, 
\end{eqnarray*}
be a filtration of $K$. 
A simplex $\sigma^i$ is called positive if it is contained in a $(k+1)$-cycle in $K^i$, and negative otherwise.
\end{definition}
We have the following useful theorem.
\begin{theorem}\label{theo:positive}~\\
Let $\sigma = \sigma^i$ be a positive $k$-simplex in the filtration of $K$. 
There exists a unique $k$-cycle $c$ that is not a boundary in $K^i$, that contains $\sigma$ and does not contain any other positive $k$-simplex. 
\end{theorem}
The $k$-cycles associated to the positive $k$-simplices in Theorem~\ref{theo:positive} allow to maintain a basis of the $k$-dim homology groups of the subcomplices of the filtration. 
At the beginning, the basis of $H_k(K^0)$ is empty, and the basis of the $H_k(K^i)$ are then built inductively. 
Then, assume that the basis of $H_k^{j-1}$ is built and the $j$-th simplex $\sigma^j$ is negative with dim of $k+1$. 
Let $c^{i_1}, \dots, c^{i_p}$ be the cycles associated to the positive simplices $\sigma^{i_1}, \dots, \sigma^{i_p}$ whose homology classes form a basis of $H_k^{j-1}$. 
Since the boundary $\partial \sigma^j$ is a $k$-cycle in $K^{j-1}$, which is not a boundary in $K^{j-1}$ but is a boundary in $K^j$, we can write
\begin{eqnarray}
  \partial \sigma^j = \sum_{k=1}^{p} \varepsilon_k c^{i_k} + b
\end{eqnarray}
where $\varepsilon_k \in \mathbb{Z}_2 = \{0,1\}$ and $b$ is a boundary of simplices with dimension $< j$. 
We then denote $\ell(j) = \max \{ i_k; \varepsilon_k = 1 \}$ and we remove the homology class of $c^\ell(j)$ from the basis of $H_k^{j-1}$. 

\begin{definition}[Persistence pair]\label{def:persistence_pair}~\\
The pairs of simplices $(\sigma^{\ell(j)}, \sigma^j)$ are called the persistence pairs of the filtration of $K$. 
\end{definition}
The homology class created by $\sigma^{\ell(j)} \in K^{\ell(j)}$ is destroyed by $\sigma^j \in K^j$. 
The persistence (duration) of this pair is, then, $j - \ell(j) -1$. 

\subsubsection{Persistence diagram}

For a fixed $k$, the persistence pairs of simplices of respective dimensions $k$ and $k+1$ are conveniently represented by a diagram on $\mathbb{R}^2$. 
Each pair $(\sigma^{\ell(j)}, \sigma^j)$ is represented by the point of coordinates $(\ell(j), j)$. 
For each positive simplex $\sigma^i$ which is not paired to any negative simplex in the filtration, we associate the pair $(\sigma^i, +\infty)$. 
Generally, if the filtration is indexed by a non-decreasing real numbers as
\begin{eqnarray}
  \emptyset = K^{\alpha_0} \subset K^{\alpha_1} \subset \dots \subset K^{\alpha_m} = K \, , \alpha_0 \leq \alpha_1 \leq \dots \leq \alpha_m \; , 
\end{eqnarray}
a persistent pair of simplices $(\sigma_i, \sigma_j)$ is represented as the point of coordinates $(\alpha_i, \alpha_j)$. 
In this case, since the sequence $\{\alpha_i\}$ is non-decreasing, some pairs can be associated to the same point on the plane. 
This can be dealt with defining a multiset. 
\begin{definition}[Multiset]\label{def:multiset}~\\
A pair of a subset $D$ of a set $S$ and a function on $D$, $m: D \to \mathbb{N}\cup \{ +\infty \}$ is called a multiset.
For $x \in D$, $m(x)$ is the multiplicity of $x$. 
\end{definition}
The support of $D$, i.e. the subset considered without the multiplicities, is denoted by $|D|$. 
Equivalently, $D$ can be represented as a disjoint union
\begin{eqnarray}
  D \equiv \bigcup_{x \in |D|} \coprod_{i=1}^{m(x)} x \; ,
\end{eqnarray}
where $\coprod$ stands for a coproduct. 
Then, we can formally define a (general) persistence diagram. 
\begin{definition}[General persistence diagram]\label{def:general_PD}~\\
Set 
\begin{eqnarray}
&&\Delta_+ \equiv \{ (x, y) \in \mathbb{R}^2 ; y \geq x \} \\
&&\Delta \equiv \{ (x, x) \in \Delta_+ ; x \in \mathbb{R}\} \; .  \end{eqnarray}
A general persistence diagram is defined as a multiset on $\Delta_+$ such that $\Delta$ is involved, the points in $\Delta_+ - \Delta$ are finite and have a finite multiplicity, and the points on $\Delta$ has infinite multiplicity.
\end{definition}
In our case, we plot a set of persistence pairs on the persistence diagram (PD). 

Since $\alpha_i < \alpha_j)$, all the points in the persistence diagram lie above the diagonal line.
By definition, points lying near the diagonal line have a short interval, indicating that they quickly die after their birth. 
In contrast, points far from the diagonal line represent the homology classes that can survive long. 
This is schematically indicated in Fig.~\ref{fig:PD_schematic}. 

\subsubsection{Bottleneck distance}

\begin{figure}
    \centering
    \includegraphics[width=0.7\textwidth]{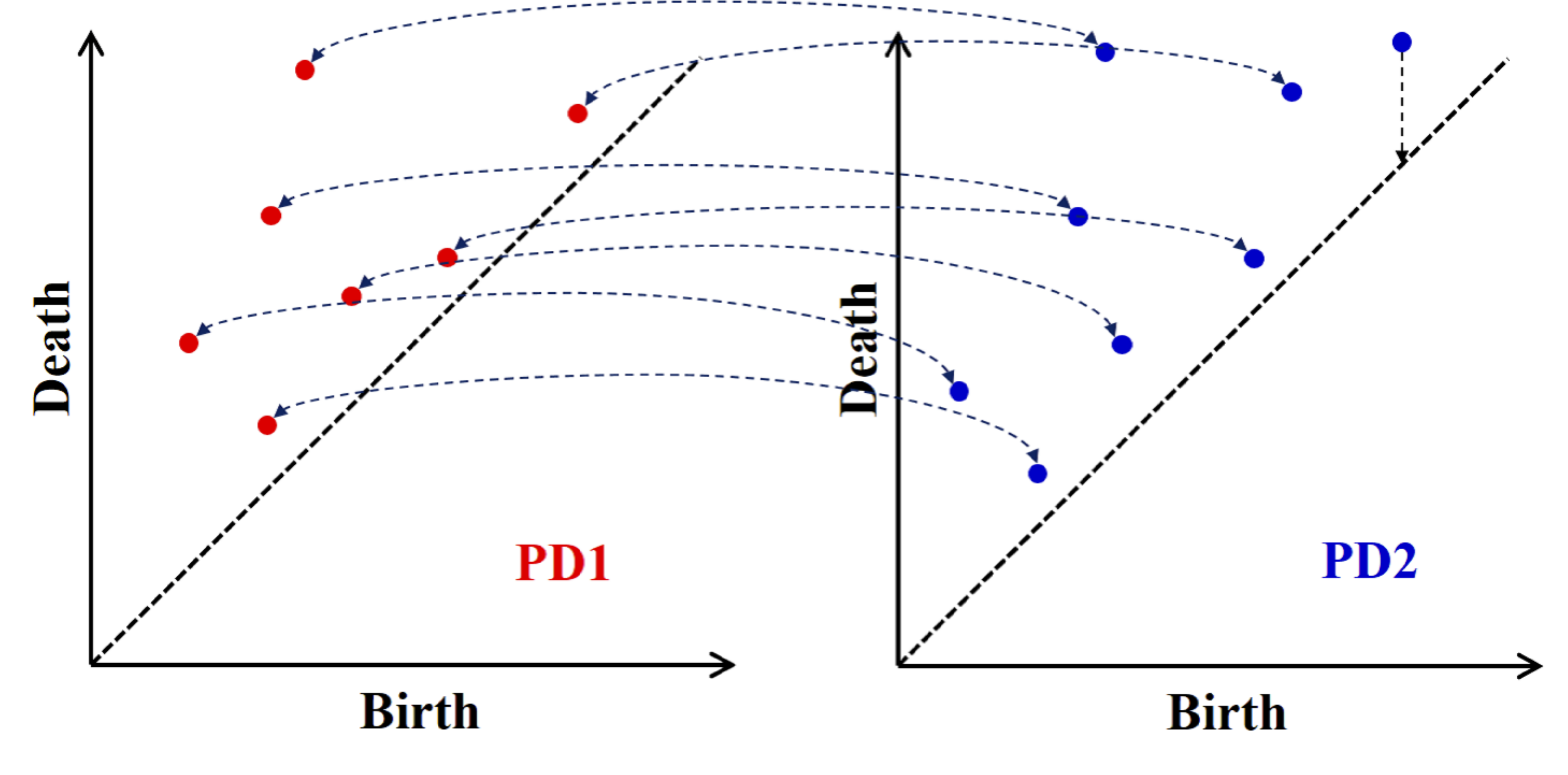}
    \caption{A schematic description how to calculate the bottleneck distance between two persistence diagrams (PDs). 
    A point without a counterpart in the other PD is regarded as corresponding to the diagonal line.
    }
    \label{fig:bottleneck_distance}
\end{figure}

We often need a method to compare the difference of two PDs for various applications. 
Since the PD is not treated a vector, we should be careful for this comparison.
The bottleneck distance is one of the appropriate measures 
\cite{DBLP:books/daglib/0025666,10.1145/3064175,cohen_steiner2007,chazal:inria-00292566}.
\begin{definition}[Bottleneck distance]\label{def:bottleneck distance}~\\
The bottleneck distance is defined as,
\begin{eqnarray}
  d_{\rm b}(D_1,D_2) \equiv \inf_{m} \sup_{x_1 \in |D_1|} \|x_1 - m(x_1)\|_{\infty}\; ,
\end{eqnarray}
where $\|x\|_{\infty}$ is a Chebyshev distance $L^{\infty} \equiv \max\{|x_1|,|x_2|\}$, $|D| \equiv D \cup \Delta$, $m:|D_1| \to |D_2|$ is a multi-bijection between two PDs and $\Delta$ is a multiset of diagonal line.
\begin{definition}[Multi-bijection]\label{def:multi_bijection}~\\
A multi-bijection between two multisets $D_1$ and $D_2$ is a bijection
\begin{eqnarray}
  \phi : \bigcup_{x_1 \in |D_1|} \coprod_{i=1}^{m(x_1)} x_1 \to \bigcup_{x_2 \in |D_2|} \coprod_{j=1}^{m(x_2)} x_2 \; .
\end{eqnarray}
\end{definition}
When two persistent diagrams are completely the same, bottleneck distance takes its minimum value.
\end{definition}
A schematic description of the bottleneck distance is presented in Fig.~\ref{fig:bottleneck_distance}. 
The stability of the bottleneck distance is guaranteed and we can safely use it for the examination of a PD (\cite{cohen_steiner2007}).

\subsection{Inverse analysis}


In the conventional BAO analysis with two point correlation function, we cannot locate the position nor specify the shape of cosmological features in a data set. 
Since this type of analysis represent an statistically features of the correlation, we can only access an average feature of the data set. 
However, these features contain essential information on the application to cosmology, or astrophysics in general. 
Therefore, solving the inverse problem from a given persistence diagram is of a vital importance for fundamental discussions on the related physics.
For a persistent set $(\ell_i, \ell_i+p_i)$ at certain time $i$ in the filtration, corresponding simplicial complexes $(\sigma_{\ell_i},\sigma_{\ell_i+p_i})$ can be obtained. 
As we have discussed in Section~\ref{subsec:PD}, the simplicial complex at the death $\ell_i+p_i$ shows the position and its shape.
In this study, we calculated $p$-values of detected holes with making use of SCHU\footnote{URL: {\tt https://github.com/xinxuyale/SCHU}}\cite{2019A&C....27...34X} along with \texttt{R TDA} package.
This enables us to specify the location and shape of the detected persistent homology classes.

We adopted this approach in this work mainly because of its conciseness, but since this field is still developing rapidly, various sophisticated methods have been proposed (see, e.g., \cite{Obayashi2018}). 
It will be meaningful to explore such method for future analysis.

\section{Performance verification by cosmological simulations}\label{sec:simulation}

\subsection{Data}

\begin{figure}
    \centering
    \includegraphics[width=0.9\textwidth]{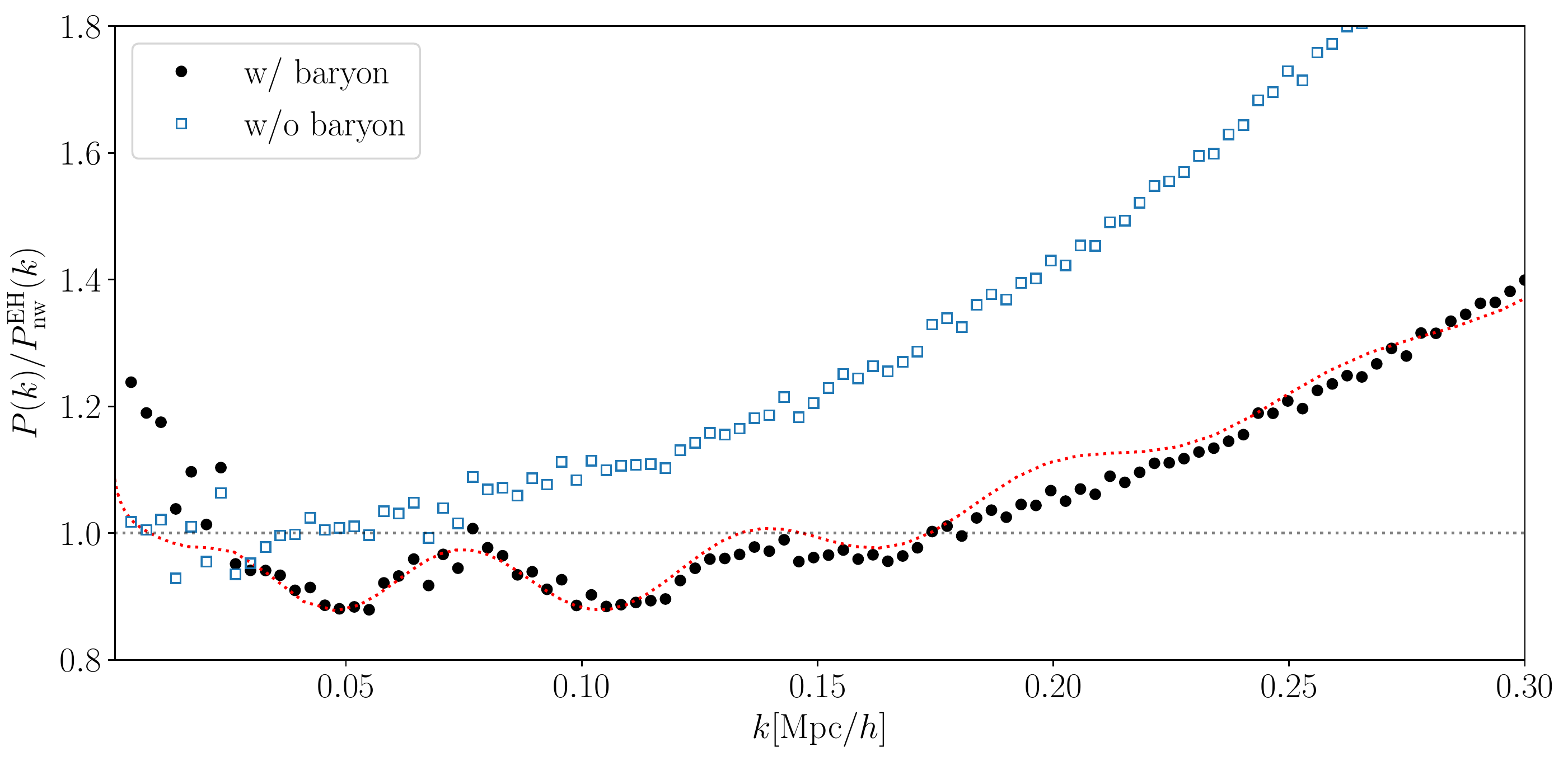}
    \caption{Measured power spectra at $z=0$ normalized by a smooth power spectrum without baryon wiggle of \cite{EisensteinHu:1998}. The sold symbol is a simulation with baryon oscillation and solid line is a halo-fit model for the same set of the cosmological parameters \citep{Takahashi+:2012}. 
    Open square is the one for without baryon oscillation.}
    \label{fig:matter_power}
\end{figure}

We performed a set of $N$-body simulations consists of only dark matter in order to accommodate a verification of our method. 
We use the publicly available $N$-body simulation code \texttt{Gadget-2} \citep{Springel:2005}, and the initial condition is produced by the second order Lagrangian perturbation theory, \texttt{2LPT} \citep{Crocce+:2006}. 
The initial condition is generated at redshift $z=20$ with the 
box size being 2~Gpc, which is sufficiently larger than the typical BAO scale of $\sim 150 \;{\rm Mpc}$. 
In order to detect the BAO from the simulation, the mass resolution is not important and thus we include only $256^3$ particles in the simulation box, which roughly corresponds to the particle mass $6.4\times 10^{13} M_{\odot}$. 
Although our simulation only include dark matter particles, the signature of BAO can be imprinted in the initial conditions since the epoch we are focusing on is well after the decoupling of photon and baryon and only we need to see is the gravitational force. 
We compute the initial power spectrum using \texttt{CLASS} \citep{Julien:2011}. 
We take a snapshot of simulation at $z=0$ and find that the BAO wiggle can be significantly detected with this data set using power spectrum analysis. 
Figure \ref{fig:matter_power} show the power spectra for our simulation set with and without baryons, normalized by smoothed power spectrum without baryon oscillation features \citep{EisensteinHu:1998}.

\begin{figure}
    \centering
    \includegraphics[width=0.9\textwidth]{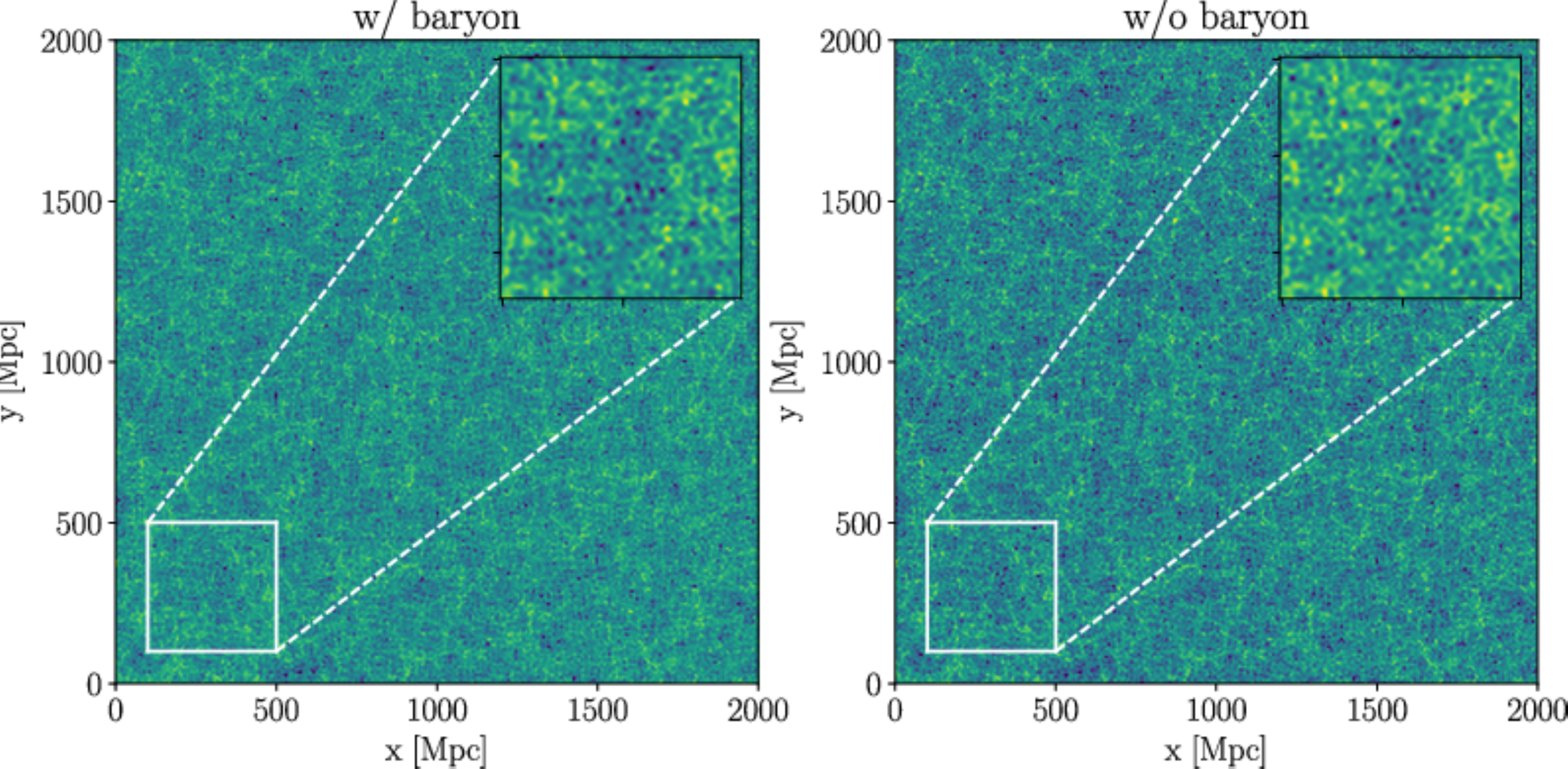}
    \caption{Simulation datasets at $z=0$ with (right) and without (left) baryon oscillation. 
    The insert is a close up of the region of 200--500~Mpc box. 
    The density fluctuations are projected along $z$-axis with 100~Mpc width, and take \texttt{arcsinh} for illustrative purpose.
    }
    \label{fig:data_simulation}
\end{figure}

As a control sample, we also generated an exactly same set of simulation but replacing the initial power spectrum without BAO wiggle. 
This can be done by reducing the amount of baryons keeping the total amount of matter (CDM + baryon) unchanged when we compute the initial power spectrum by \texttt{CLASS}. 
As expected, the measured power spectrum does not exhibit the BAO wiggle for this control sample.
The assumed baryon density parameter for each case is $\Omega_{B} = 0.049$ (w/ baryon) and $\Omega_{\rm B} = 0.002204$ (w/o baryon), respectively. 
Figure~\ref{fig:data_simulation} shows the snapshot of our simulation with and without baryon oscillation feature. 
Since we use the same random seed for both simulation set, it is fairly difficult to see the difference between them by a visual inspection. 

In this paper, for the TDA analysis, we randomly extracted 2000 particles from the parent sample of $256^3$ particles simply because 
of the computational limitation in our system. 
In our future works we will upgrade our algorithm to handle much larger data set by parallelize the code or use of more powerful resources.

\subsection{Result of the simulation analysis}

We discuss the PDs from the simulations. 
In order to evaluate the statistical significance of the detected signals, we need a method to obtain the confidence intervals.
One of the ways to determine the confidence intervals on a PD is to use the bottleneck distance. 
In this study, we measure the distance between a pair of persistent diagrams obtained from bootstrap resampling samples by the bottleneck distance.

\begin{figure}
    \centering
    \includegraphics[width=0.7\textwidth]{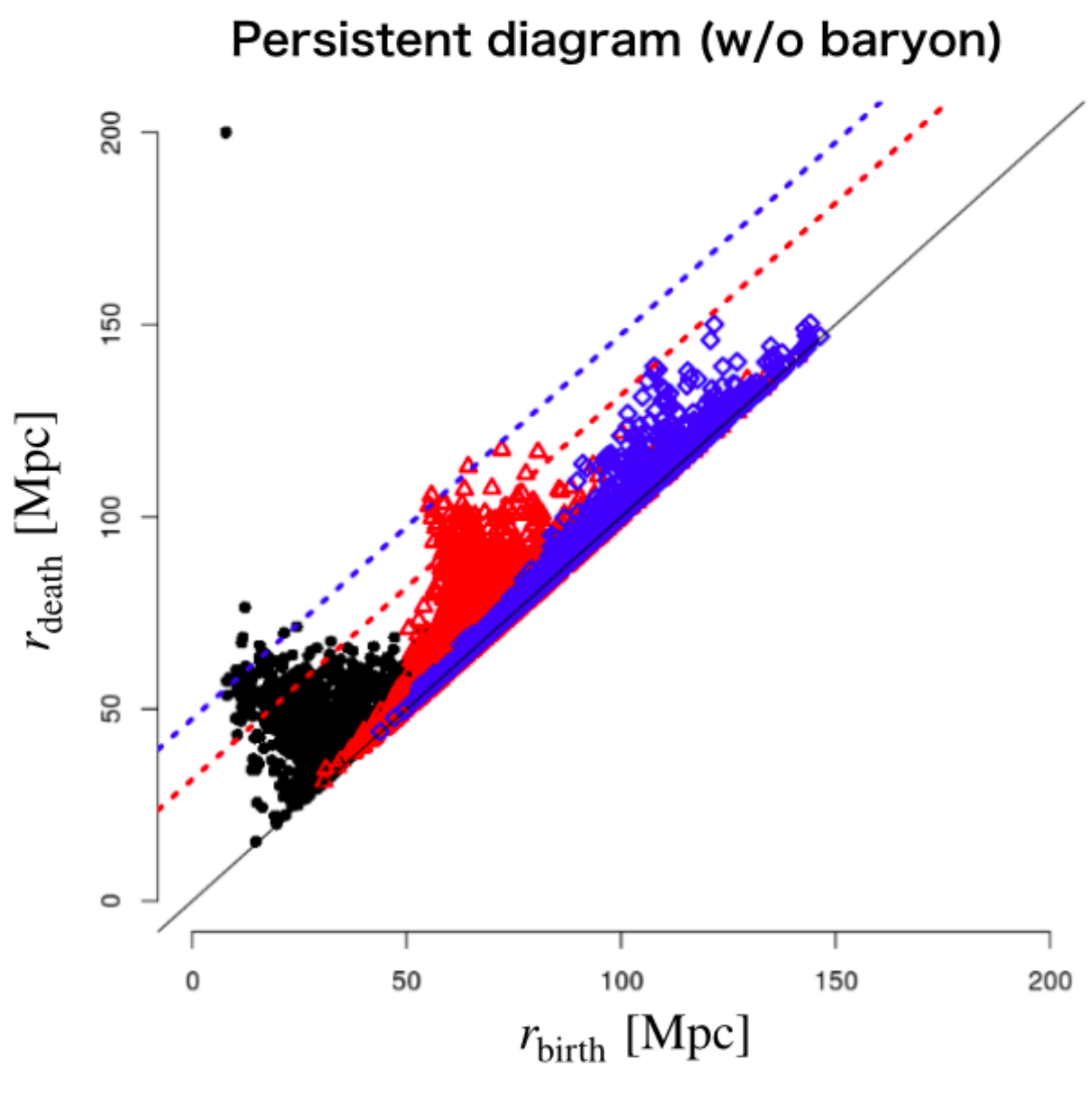}
    \includegraphics[width=0.7\textwidth]{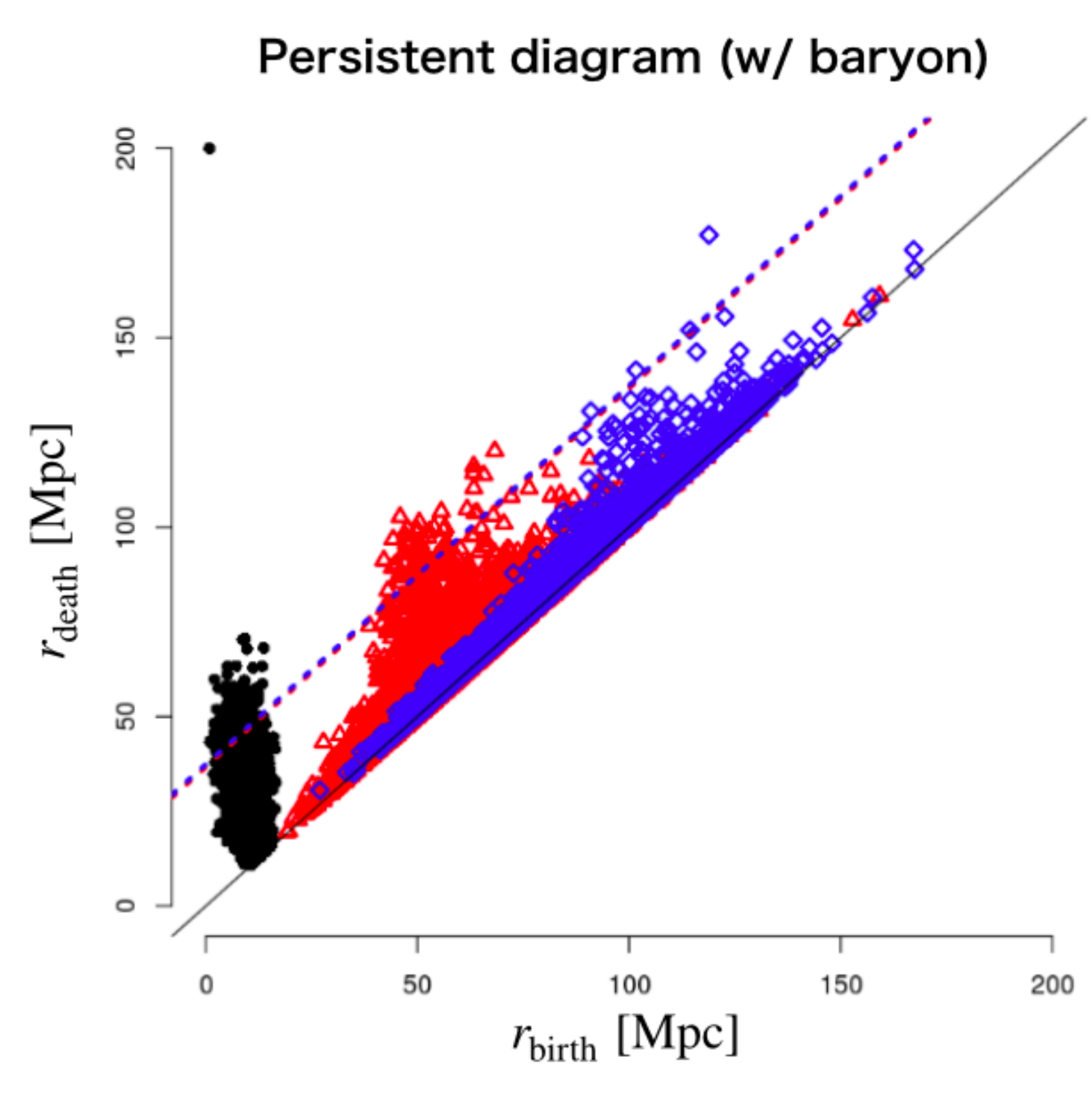}
    \caption{Persistent diagrams (PDs) of the simulation datasets. Black dots, red triangles and blue diamonds are $H_0$, $H_1$ and $H_2$ respectively. Diagonal black solid line is $r_{\rm birth}=r_{\rm death}$.
    Upper: PD of the simulation without baryon effect, Lower: PD of the simulation with baryon effect. 
    }
    \label{fig:PD_simulation}
\end{figure}

The obtained PDs for the simulations are shown in Fig.~\ref{fig:PD_simulation}. The upper panel is for (w/o) baryon and the lower panel is for (w/) baryon. The black points, red triangles and blue diamonds are $H_0$, $H_1$ and $H_2$ points.
The red and blue dotted diagonal lines are 90~\% confidence bands for $H_1$ and $H_2$, respectively. 
These confidence bands for this PD were calculated by bootstrap resampling
with $N_{\rm boot}=30$. 
In this study, we regard points lying above the corresponding confidence bands as being statistically significant.

\begin{figure}
    \centering
    \includegraphics[width=0.9\textwidth]{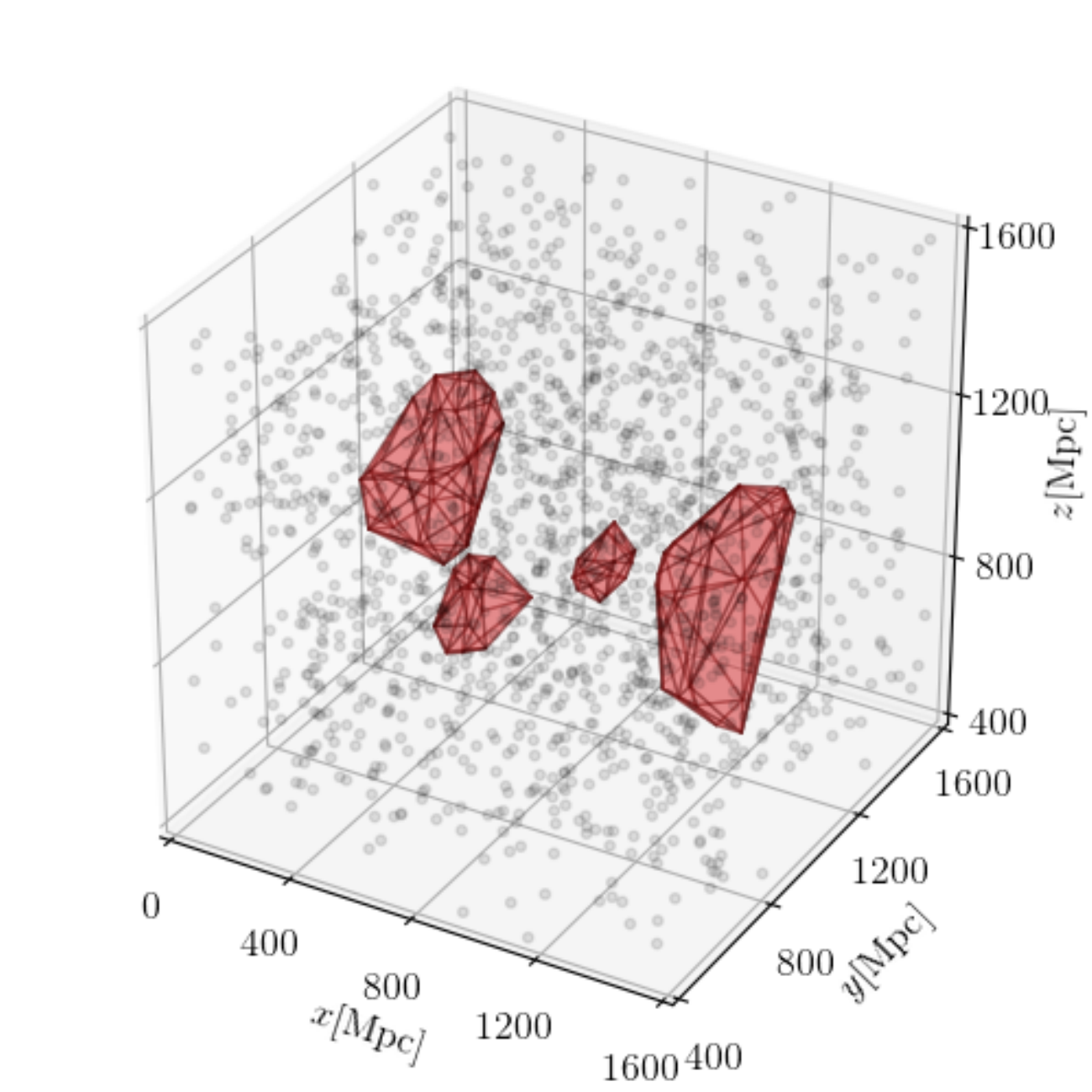}
    \caption{
    An example of inverse analysis. We show the real space structure for 4 highest significant $H_2$ in the simulated data. The red shaded region is a convex hull of all the constituent particles.
    }
    \label{fig:inverse_analysis_simulation}
\end{figure}

As can be seen in the bottom panel of Fig.~\ref{fig:PD_simulation}, we detected four significant $H_2$, $3\mbox{-}$dimensional holes (blue diamonds) in the (w/) baryon sample. However, there is no significant hole in the (w/o) baryon sample. The $r_{\rm birth}$ and $r_{\rm death}$ for significant holes detected for (w/) baryon sample are displayed in Tab.~\ref{tab:PD_simulaiton}. The mean $r_{\rm death}$ is $150.16\pm 8.46\ [{\rm Mpc}]$. Although the number of detected hole is only four, obtained $\bar{r}_{\rm death}$ is consistent with the radius that is expected for BAO signal. The position an the shape of detected $H_2$ is displayed in Fig.~\ref{fig:inverse_analysis_simulation}.

For $H_1$ homology (red triangles), we obtained significant features from both of the sample. We detected 17 significant $H_1$ homology whose $p\mbox{-}$value is less than $0.2$ from the simulation data with baryon. The mean $r_{\rm death}$ for $H_1$ homology is $99.00\pm 2.26\ {\rm [Mpc]}$. 
For the sample without baryon, we detected 34 $H_1$ homology and $\bar{r}_{\rm death}=100.49\pm 3.24\ {\rm [Mpc]}$. The agreement in the characteristic scale in $H_1$ suggests that there are loop-like structures whose scale is not affected by the existence of baryon in $2\mbox{-}$dimensional space. Cosmic filament can be a candidate for these structure. Meanwhile, there is a difference in $\bar{r}_{\rm birth}$ between the simulation setup. 
The mean birth radius is $46.13\pm 2.24\ [{\rm Mpc}]$ for (w/) baryon sample while $\bar{r}_{\rm birth}= 62.34\pm 2.84\ [{\rm Mpc}]$ for (w/o) baryon sample. 
From this analysis, the separation between galaxies that construct loops become smaller with baryon.

For $H_0$ homology (black points), we found clear difference in distribution of PH between (w/) and (w/o) baryon sample. 
The $H_0$ corresponds to sequential structure, such as galaxy clusters and filaments. 
For the sample (w/o) baryon, $r_{\rm birth}$ distributes almost uniformly along the diagonal line. On the other hand, the distribution of $r_{\rm birth}$ concentrates to low value for (w/) baryon sample. This means the separation between two points in this sample is roughly the same.
As for $H_0$ and $H_1$, further analysis is clearly needed to interpret what we discovered from the PD. 
This is left for our subsequent future works. 

\begin{table}\label{tab:PD_simulaiton}
\begin{center}
  \begin{tabular}{ccc} \hline\hline
    No. & $r_{\rm birth}\ {\rm [Mpc]}$ & $r_{\rm death}\ {\rm [Mpc]}$\\ \hline
    1 & 116.34 & 130.65  \\
    2 & 127.42 & 141.80  \\
    3 & 140.17 & 151.70  \\
    4 & 144.60 & 176.47  \\ \hline
  \end{tabular}
  \caption{The $r_{\rm birth}$ and $r_{\rm death}$ for holes detected in our simulated data whose $p\mbox{-}$value is less than 0.2.}
\end{center}
\end{table}

\section{Analysis with Sloan Digital Sky Survey Data Release 14}\label{sec:SDSS}

\subsection{Data}

We this study we used quasar data from the extended
Baryon Oscillation Spectroscopic Survey (eBOSS) \cite{2016AJ....151...44D} LSS catalog \cite{2018MNRAS.473.4773A} in the SDSS DR14 \cite{2018A&A...613A..51P}.
This is a part of the SDSS-IV \cite{2017AJ....154...28B}.   
The observed bands are $u$, $g$, $r$, $i$, and $z$. 
The sky coverage area is $2044\ \deg^2$ and covers a redshift range $0.8<z<2.2$. 
The effective area is $1288\ {\deg}^2$ the Northern Galactic Cap (NGC) and $995\ {\deg}^2$ Southern Galactic Cap (SGC). 
The quasar selection was done in \cite{2015ApJS..221...27M}. 
We used the data The BAO signal in this sample is already examined and confirmed in the literature \cite{2018MNRAS.473.4773A}.

Although the whole sample contains 147,000 quasars, we randomly sampled 2000 galaxies.
Namely, we do not use the full dataset but a sparsely drawn subsample of the SDSS red galaxies. 

\subsection{Result}

\begin{figure}
    \centering
    \includegraphics[width=0.7\textwidth]{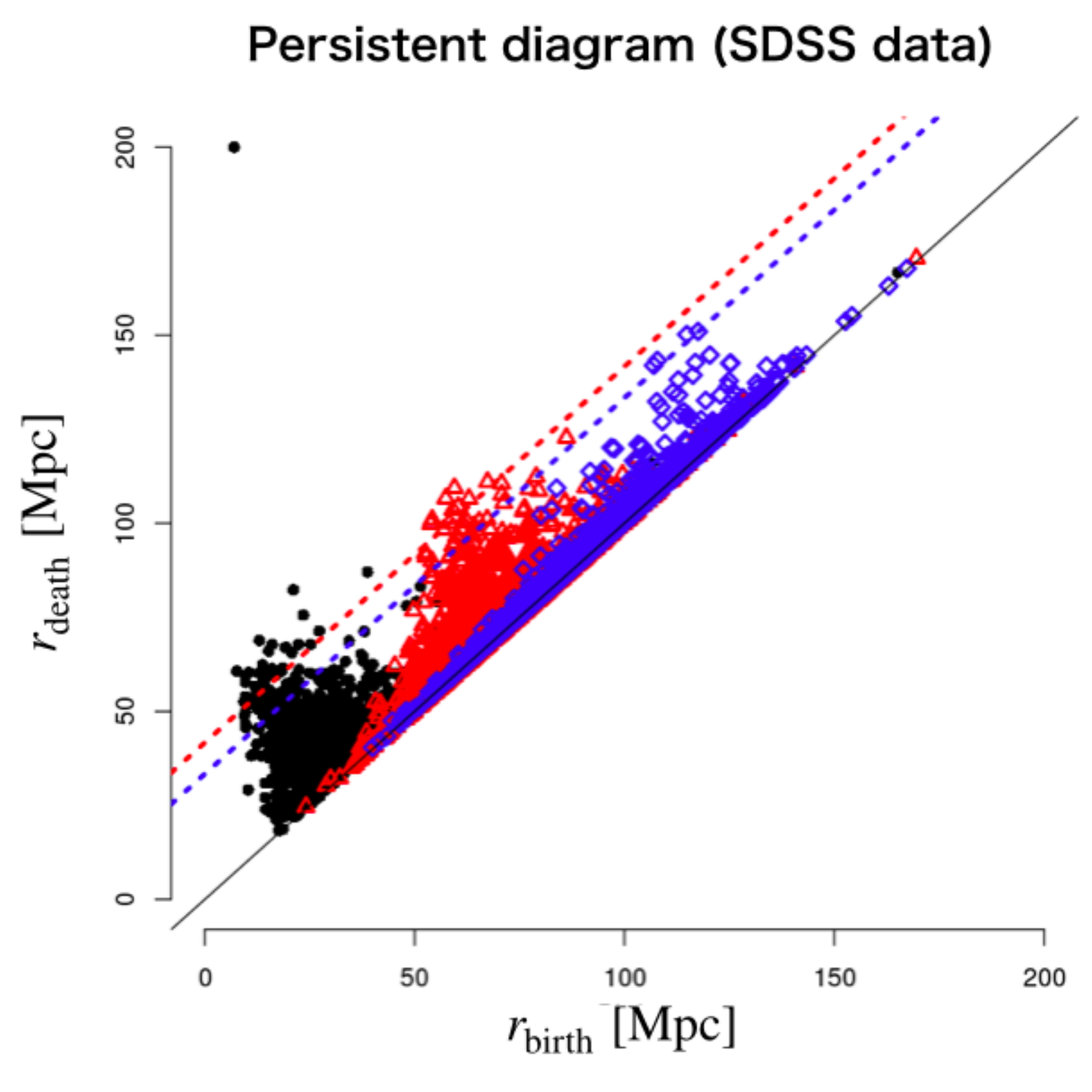}
    \caption{Persistent diagrams (PDs) of the SDSS data. 
    }
    \label{fig:SDSS}
\end{figure}

\begin{figure}
    \centering
    \includegraphics[width=0.9\textwidth]{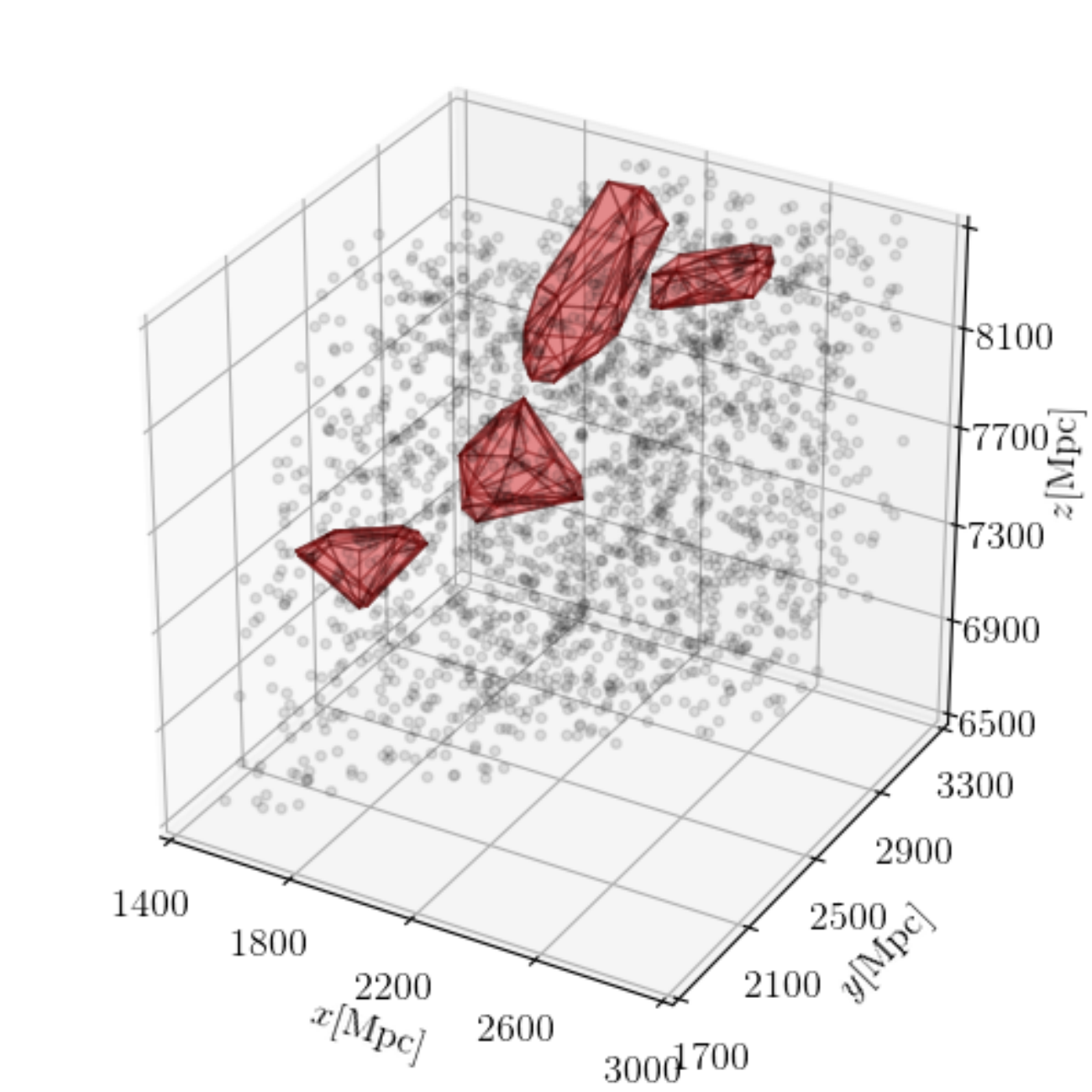}
    \caption{
    An example of inverse analysis. We show the real space structure for 4 highest significant $H_2$ in the simulated data. The red shaded region is a convex hull of all the constituent particles.
    }
    \label{fig:inverse_analysis_SDSS}
\end{figure}

The PD for SDSS DR14 data is displayed in Fig.~\ref{fig:SDSS}. We obtained four significant $H_2$ homology, "shells", from SDSS DR14 data. The mean $r_{\rm death}$ is, $146.6\pm 2.0\ {\rm [Mpc]}$. Although the number of detected $H_2$ homology is low, the $\bar{r}_{\rm birth}$ is consistent with the expected value for BAO signal. The result of inverse analysis for $H_2$ is displayed in Fig.\ref{fig:inverse_analysis_SDSS}. In this analysis, 19 $H_1$ homology are detected as significant loop. The mean $r_{\rm death}$ for $H_1$ homology is $101.82\pm 3.54\ {\rm [Mpc]}$. This radius agreed with the $\bar{r}_{\rm birth}$ which we have obtained from the simulation data. The birth radius for the data is $57.92\pm3.05\ {\rm [Mpc]}$. Comparing with the results from simulation data, this value agrees with $\bar{r}_{\rm birth}$ without baryon. The distribution of $H_0$ homology is similar to that of (w/o) baryon sample from simulation. This means the separations between two galaxies takes some range in their values. We need to note that we did not detected any significant $H_0$ from the data. Therefore we can only discuss about the trend of distribution in PD.

\begin{table}\label{tab:PD_data}
\begin{center}
  \begin{tabular}{ccc} \hline\hline
    No. & $r_{\rm birth}\ {\rm [Mpc]}$ & $r_{\rm death}\ {\rm [Mpc]}$\\ \hline
    1 & 107.00 & 141.98  \\
    2 & 107.83 & 143.30  \\
    3 & 114.86 & 150.20  \\
    4 & 117.56 & 150.98  \\ \hline
  \end{tabular}
  \caption{The $r_{\rm birth}$ and $r_{\rm death}$ for holes detected in SDSS DR14 data whose $p\mbox{-}$value is less than 0.2.}
\end{center}
\end{table}
\section{Summary and discussion}\label{sec:summary}
In this study, we verified the usefulness of TDA on detecting BAO signal from galaxy distribution. PH constructs filtration defined by simplicial complex for given scale and dimension. This property is particularly useful when we analyse the scale and shape of holes.

We examined the effect of baryon on PHs with $N$-body simulation with $N=256^3$. We limited the number of particles to 2,000 due to the limitation of our system. Simulations with and without baryon physics showed drastically different PHs. 
Not only the 2-dim shells but also the behavior of 0- and 1-dim loops was found to be quantitatively different. 
Although we detected four significant $H_2$ homology from (w/) baryon sample, no $H_2$ was detected from (w/o) baryon sample. We also found clear difference in the distribution of $H_1$ and $H_0$ homology. While the $\bar{r}_{\rm death}$ of $H_1$ has equivalent value, $\bar{r}_{\rm birth}$ significantly differs. Similarly, $\bar{r}_{\rm birth}$ for $H_0$ has different trend between (w/) and (w/o) baryon sample.
Obviously, we need to increase the number of particles that is used in the PD analysis. This lack of particle will cause to miss accurate structure in the data set.

We succeeded in detecting the BAO signal only with a 2000 subsample of quasars from SDSS. 
The obtained $\bar{r}_{\rm death}$ is $146.6\pm 2.0\ {\rm [Mpc]}$ for $H_2$ which is consistent with the scale of BAO signal.
This means that the PH is computationally far much less expensive than the correlation function method. 
Clearly the PH approach will shed light to the cosmological analysis of BAOs. 

Our successful detection of the BAO makes a clear contrast to the traditional 2-point correlation function method.
The 2-point correlation requires a very large and dense-sampled galaxy data, while the PH can detect the BAO signal only with sparse-sampled data of 2000 galaxies. 
This means that, with the PH analysis, a sparse sampled galaxy survey can be used for cosmological studies. 
It will bring a new insight for the design and strategy of next generation cosmological galaxy surveys. 

Further, we can visualize the BAO structure by the inverse analysis of the PH. 
We stress that this particular function of the inverse analysis in the TDA is worthy of close attention, indeed a distinctly different feature of the method. 
Since we can straightforwardly specify the structure contributing to the BAO signal, it makes various cosmological discussions much easier than previous analysis. 
More detailed results and cosmological tests are left to be presented in our subsequent works.

\acknowledgments
We thank Shiro Ikeda, Kenji Fukumizu, Satoshi Kuriki, and Yoh-ichi Mototake for fruitful discussions and suggestions. 
This work has been supported by JSPS Grants-in-Aid for Scientific Research (17H01110 and 19H05076), and Grant-in-Aid for Scientific Research on Innovative Areas “Cosmic Acceleration” (15H05890 and 16H01096). 
This work has also been supported in part by the Sumitomo Foundation Fiscal 2018 Grant for Basic Science Research Projects (180923), and the Collaboration Funding of the Institute of Statistical Mathematics ``New Development of the Studies on Galaxy Evolution with a Method of Data Science''.

\appendix

\renewcommand{\thedefinition}{\Alph{section}\arabic{definition}}
\renewcommand{\thetheorem}{\Alph{section}\arabic{theorem}}

\section{Algebraic preliminaries}
\setcounter{definition}{0}
\setcounter{theorem}{0}

We introduce basic terminologies to understand homotopy in algebra. 

\begin{definition}[Equivalence relation]\label{def:equivalence_relation}
~\\
A given binary relation $\sim$ on a set $X$ is an equivalence relation if and only if it is reflexive, symmetric and transitive. 
That is, for all $a$, $b$ and $c \in X$, 
\begin{enumerate}
    \item Reflexivity: $a \sim a$, 
    \item Symmetry: $a \sim b$ if and only if $b \sim a$, 
    \item Transitivity: if $a \sim b$ and $b \sim c$ then $a \sim c$.
\end{enumerate}
\end{definition}

\begin{definition}[Contractible]\label{def:contractible}~\\
When a set $X$ is homotopy equivalent to a one-point space, $X$ is contractible.
\end{definition}

\begin{definition}[Homotopy]~\\
For continuous maps $f, g : X \rightarrow Y$, a continuous map $F$ from a product space of a closed interval $I = [0,1]$ and $X$ to $Y$
satisfying
\begin{eqnarray}
  F_{X\times\{0\}} = f, \quad F_{X\times \{1\}} = g
  \end{eqnarray}
exists, $f$ and $g$ are said to be homotopic, denoted as $f \simeq g$.
Here
\begin{eqnarray*}
  F_{X\times\{t\}}(x) \equiv F(x, t)
\end{eqnarray*}
$F$ is referred to as the homotopy from $f$ to $g$. 
\end{definition}
It follows that two homotopic maps $f$ and $g$ can be transformed continuously to each other by changing $t$ via $F$.
Then the binary relation $\simeq$ is an equivalence relation (see Definition~\ref{def:equivalence_relation}). 

\begin{definition}[Homotopy equivalence]
A continuous map $f : X \rightarrow Y$ is a homotopy equivalence if there exist a continuous map $g : Y \rightarrow X$ such that 
\begin{eqnarray}
  g \circ f \simeq 1_X \mbox{ and } f \circ g \simeq 1_Y \; , 
\end{eqnarray}
where $1_X$ and $1_Y$ are identity maps on $X$ and $Y$, respectively. 
If a homotopy equivalence exits, $X$ and $Y$ are homotopy equivalent, denoted as $X \simeq Y$. 
\end{definition}
Intuitively, it means that $X$ and $Y$ are homotopy equivalent if they can be transformed into each other by bending, shrinking and expanding. 
The homotopy equivalence is fundamentally important because many concepts in algebraic topology are homotopy invariant, that is, they respect the relation of homotopy equivalence. 
Particularly in the TDA, the homology group is homotopy invariant. 
The homology group of a simplicial complex is compatible with computers and easy to calculate. 
Then, even if a homology group of a certain shape is difficult to calculate directly, we can calculate it by constructing the homotopy equivalent simplicial complex. 
This guarantees all the analysis presented in this article.

\bibliographystyle{JHEP}
\bibliography{Kono2020b}

\end{document}